%% file: main.tex
\documentclass[twocolumn]{aastex62}

\usepackage{longtable}
\usepackage{graphicx}
\usepackage{amsmath, amssymb}
\usepackage{placeins}
\usepackage{booktabs}
\usepackage{color}
\usepackage{units}
\usepackage{epstopdf}
\usepackage[caption=false]{subfig}
\usepackage{float}
\usepackage[caption=false]{subfig}

\newcommand{\ffrac}[2]{\ensuremath{\frac{\displaystyle #1}
                                        {\displaystyle #2}}}
\newcommand{\Nrec}{N_{\text{rec}}}
\newcommand{\Ntot}{N_{\text{tot}}}

\newcommand{\lambdasne}{\Lambda_{\text{SNe}}}
\newcommand{\psne}{p_{\text{SNe}}}
\newcommand{\blambda}{\pmb{\lambda}}
\newcommand{\bgamma}{\pmb{\gamma}}
\newcommand{\bkappa}{\pmb{\kappa}}
\newcommand{\bbeta}{\pmb{\beta}}
\newcommand{\mstar}{m_{*}}
\newcommand{\mstamp}{m_{\text{inj}}}
\newcommand{\mclst}{\texttt{CLASS\_STAR}}
\newcommand{\mfwhm}{\texttt{FWHM}}
\newcommand{\mellip}{\texttt{ELLIP}}
\newcommand{\rb}{\texttt{RB2}}
\newcommand{\thetaiq}{\Theta_{\text{IQ}}}
\newcommand{\fsky}{F_{\text{sky}}}
\newcommand{\shost}{S_{\text{gal}}}
\newcommand{\phiiq}{\Phi_{\text{IQ}}}
\newcommand{\mlim}{m_{\text{lim}}}
\newcommand{\epsilonhat}{\hat{\varepsilon}}

\begin{document}
\title{Towards Rate Estimation For Transient Surveys I:

\small{
Assessing transient detectability and
volume sensitivity for iPTF
}
}

\author[0000-0003-0038-5468]{Deep Chatterjee}
\affiliation{Department of Physics, University of Wisconsin, Milwaukee, WI 53201, USA}

\author[0000-0002-3389-0586]{Peter E. Nugent}
\affiliation{Department of Astronomy, University of California, Berkeley, CA 94720-3411, USA}
\affiliation{Lawrence Berkeley National Laboratory, Berkeley, CA 94720, USA}

\author[0000-0002-4611-9387]{Patrick R. Brady}
\affiliation{Department of Physics, University of Wisconsin, Milwaukee, WI 53201, USA}

\author[0000-0003-2667-7290]{Chris Cannella}
\affiliation{Cahill Centre for Astrophysics, California Institute of Technology,
             1200 East California Boulevard, Pasadena, CA 91125, USA}

\author[0000-0001-6295-2881]{David L. Kaplan}
\affiliation{Department of Physics, University of Wisconsin, Milwaukee, WI 53201, USA}

\author[0000-0002-5619-4938]{Mansi M. Kasliwal}
\affiliation{Cahill Centre for Astrophysics, California Institute of Technology,
             1200 East California Boulevard, Pasadena, CA 91125, USA}

\begin{abstract}
\input{abstract}
\end{abstract}

\section{Introduction}
\label{sec:intro}
\input{introduction}

\section{Intermediate Palomar Transient Factory}
\label{sec:isp}
\input{isp}

\section{Fake Transients}
\label{sec:injections}
\input{injections}

\section{Single Epoch Detectability}
\label{sec:efficiency}
\input{efficiency}

\section{Lightcurve Recovery}
\label{sec:rates}
\input{rates}

\section{Discussion and Conclusions}
\label{sec:discussion}
\input{discussion}

\acknowledgments
\input{acknowledgements}

\appendix
\input{appendix}
\bibliography{references}

\end{document}

%% file: abstract.tex
The last couple of decades have seen an emergence of transient detection
facilities in various avenues of time domain astronomy which has provided us
with a rich dataset of transients. The rates of these transients have
implications in star formation, progenitor models, evolution channels
and cosmology measurements.
The crucial component of any rate calculation is the detectability and
space-time volume sensitivity of a survey to a particular transient type as
a function of many intrinsic and extrinsic parameters. Fully sampling that
multi-dimensional parameter space is challenging.  Instead, we present
a scheme to assess the detectability of transients using supervised
machine learning. The data product is a classifier that determines the detection
likelihood of sources resulting from  an image subtraction pipeline associated
with time domain survey telescopes, taking into consideration the intrinsic
properties of the transients and the observing conditions. We apply our method
to assess the space-time volume sensitivity of type Ia supernovae (SNe~Ia) in
the intermediate Palomar Transient Factory (iPTF) and obtain the result,
$\langle VT \rangle_{\mathrm{Ia}} = 2.93 \pm 0.21 \times 10^{-2}
\; \mathrm{Gpc^{3}\,yr}$.
With rate estimates in the literature, this volume sensitivity gives
a count of $680-1160$ SNe~Ia detectable by iPTF which is consistent
with the archival data.  With a view toward wider applicability of this
technique we do a preliminary computation for long-duration type IIp
supernovae (SNe~IIp) and find
$\langle VT \rangle_{\mathrm{IIp}} = 7.80 \pm 0.76 \times 10^{-4}
\; \mathrm{Gpc^{3}\,yr}$.
This classifier  can be used for computationally fast space-time volume sensitivity
calculation of any generic transient type using their lightcurve properties. Hence,
it can be used as a tool to facilitate calculation of transient rates in a range of
time-domain surveys, given suitable training sets.

%% file: introduction.tex
The last two decades have brought about a revolution in the field of time-domain optical
astronomy with experiments like 
\replaced{
Pan-STARRS, \citep{panstarrs1} Sloan Digital
Sky Survey, \citep{sdss} the ATLAS survey, \citep{atlas} the Catalina survey, \citep{catalina}
the All-Sky Automated Survey for Supernovae, \citep{2019MNRAS.484.1899H}
the Palomar and intermediate Transient Factory (PTF), \citep{ptf} and Zwicky Transient
Facility (ZTF) \citep{ztf_kulkarni}
}
{
Sloan Digital Sky Survey, \citep{sdss} the Palomar and intermediate Transient Factory
(PTF), \citep{ptf} the Catalina survey, \citep{catalina} Pan-STARRS, \citep{panstarrs1}
the ATLAS survey, \citep{atlas} Zwicky Transient Facility (ZTF) \citep{ztf_kulkarni}
and the All-Sky Automated Survey for Supernovae, \citep{2019MNRAS.484.1899H}
}
performing all sky searches with rolling cadence to
locate transients.
The timescale of these transients varies from a few minutes, like M
dwarf flares, up to a few weeks or months, like supernovae.

Studying transient rates is essential to understand the progenitor systems and
environments they occur in. For example, while core-collapse supernovae are associated
with more recent massive stars, type Ia supernovae occur in both younger and older
populations \citep{2012PASA...29..447M}. The distribution of transients in space and time
helps us understand metal enrichment, galaxy formation and the overall evolution
of the universe. The classification and compilation of transients from the surveys
provide a rich dataset which can be used to make statements about their rates
and population. Next generation surveys like the Large Synoptic Survey Telescope
\citep{lsst} are expected to make significant additions to already existing catalogs
with wide-deep-fast searches.

A quantitative assessment of the transient detectability by the survey is an essential
component required to study transient rates. A survey could miss the
observation and confirmation of transients for reasons of being intrinsically dim,
occurring when the instrument was not observing, poor weather conditions and so on.
Therefore, it is crucial to understand the circumstances under which the
survey is sensitive in recovering transients. The transient detectability leads
to the calculation of a space-time sensitive volume to particular
transient types. 
This depends on properties of the source and its environment, like its
brightness or its host galaxy brightness. The instrument cadence and observing
schedule are also expected to contribute significantly.
A fast cadence is necessary to capture the evolution of, say, an M dwarf flare
which last a few minutes, as opposed to a supernova, which evolves for a couple of
months.

We consider the intermediate Palomar Transient Factory (iPTF), the successor
of PTF and predecessor of ZTF. As a first step, we assess the efficiency of the
real-time image subtraction pipeline.
We insert fake transients with varying properties into the
original iPTF images and then run the pipeline to test recovery.
This forms our \emph{single-epoch} detectability.
While this step is similar to the work done for the PTF pipeline by \cite{frohmaier_2017},
our analysis differs in final data product for the single-epoch detectability.
We make use of supervised machine learning to train a classifier on missed and
found fake transients reported by the pipeline to make predictions about the
detectability of an arbitrary transient. For completeness, we note that the
performance of the survey in the galactic plane is expected to be different
from the high latitude fields and requires a separate analysis. The analysis
presented in this paper could be applied to only galactic fields to obtain the
detection efficiency in the galactic plane. Here, we study the
detectability in the high latitude fields or, alternatively, of transients of
extra galactic origin.
Under such a consideration, this step is independent of the transient type. 
The multi-epoch observation and detection of a transient can be done using the
single-epoch detectability at each epoch.
The use of machine learning in this case has advantages in the 
areas of computing time, determination of systematic errors, ease of improving accuracy
at the cost of computing time when required, and handling correlation between training parameters. 
As a second step, we consider the transient lightcurve evolution. We simulate transient
lightcurves in space-time and use the iPTF observing schedule in conjunction with
this classifier to get the epochs at which the transient is detected. We restrict
to type Ia and type IIp supernova lightcurves in this work, the former being the primary
result. For the type Ia supernovae (SNe~Ia), we impose a minimum number of five epochs
of detection brighter than 20th magnitude with at least two during the rise and at least two 
during the fall of the lightcurve to be a ``confirmed'' SN~Ia. The simulated
SNe~Ia are used to do a Monte-Carlo integral over space-time to obtain the space-time
volume sensitivity. For the type IIp supernovae (SNe~IIp) lightcurves, the procedure
is the same, except we consider a IIp lightcurve recovered if there are at least
five epoch observations brighter than 20th magnitude within a span of three weeks
during the ``plateau'' phase.

The organization of the paper is as follows. In Sec.~\ref{sec:isp} we give a
brief description of the iPTF real-time image subtraction pipeline.
In Sec.~\ref{sec:injections} we give details of the procedure of injecting fake
transients into original iPTF images. We present the results after running
the image subtraction pipeline in Sec.~\ref{sec:efficiency}. Here, we select a
subset of parameters that captures maximum variability in detecting transients,
train a classifier based on the missed and found fake transients and cross validate
the performance of the classifier. In Sec.~\ref{sec:rates} we use a SN~Ia
lightcurve model to simulate an ensemble of transients uniform in co-moving
volume, pass them through the four year observing schedule and determine
the fraction which would be detectable by iPTF. This is then used to compute
the space-time volume sensitivity for SNe~Ia. A similar but simpler analysis
is also done for SNe~IIp to obtain its space-time sensitive volume.
Finally, in Sec.~\ref{sec:discussion} we present the procedure of getting the
rate posterior assuming the detections to be a Poisson process with a
mean intrinsic rate.

%% file: isp.tex
The intermediate Palomar Transient Factory (iPTF) was a survey operated
at the Palomar Observatory between late 2012 and early 2017. It had two filters:
$R$ (centered at $6581$ \AA) and $g$ (centered at $4754$ \AA).
It performed fast-cadence experiments resulting in about $300-400$
exposures on a good night with a nightly output of about $50-70$ GB.
The images were processed by the real-time image subtraction pipeline
to report transients within minutes latency. Details are presented in
\cite{nugent} and \cite{cao2016}. Here, we give a brief description.

\subsection{iPTF Image Subtraction Pipeline}

The iPTF real-time image subtraction pipeline (henceforth ISP) was
hosted at the National Energy Research Scientific Computing Center (NERSC).
A complete exposure of 11 working CCDs was transferred to NERSC
immediately after data acquisition to search for new candidates.
The pipeline preprocessed the images to remove bias
and correct for flat-fielding. It solved for astrometry and
photometry, and performed image subtraction using the \texttt{HOTPANTS}
algorithm \citep{2015ascl.soft04004B}. New candidates were assigned a
\emph{real-bogus} classification score between 0 and 1 corresponding
to bogus and real respectively \citep{real_bogus_2}. Additionally,
candidates would be cross-matched to external catalogs to remove
asteroids, active galactic nuclei (AGNs) and variable stars.

%% file: injections.tex
In order to quantify the performance of the iPTF
ISP, we perform an end to end simulation
using fake transients. We inject fake point source
transients in the iPTF images and then run the
pipeline on both the original images and the faked ones.
The transients are either missed or found
by the ISP, which forms the detectability.
We find the efficiency by binning up the parameter space
and taking ratio of found to total transients in them.
Regarding the mnemonic in subsequent sections, we make a
distinction between the terms \emph{detectability} and \emph{efficiency}.
Detectability is a decision taken in the sense of a yes/no, while,
efficiency is the ratio mentioned above.
The former is a binary decision, either of $\{0, 1\}$,  while the latter
is a quantity $\in [0, 1]$.

\subsection{Point Source Transients}\label{sec:point_source_transients}
We follow the \emph{clone stamping} technique used by \cite{frohmaier_2017}
for PTF to perform our fake point source injections. The parameters
describing these fake transients are \emph{single epoch} - they represent the
intrinsic properties of the object and observing conditions at a
particular epoch. In other words, here we assess the detectability
given the transient was in the field of view of the instrument.

The computational cost for performing injections into all iPTF images
and running ISP on them is significant. Therefore, we carry out the
process in a single iPTF field 100019. We choose this field since the distribution
of transient population in this field is an accurate representation of the
transient population in the sky observed from Palomar (see Fig. 1 of
\cite{frohmaier_2017}).

\deleted{subsubtitle \textbf{Injection Procedure}}

\begin{figure*}[htp]
\begin{center}
    \subfloat[Brighter fake transient]{%
        \includegraphics[width=0.32\textwidth]
        {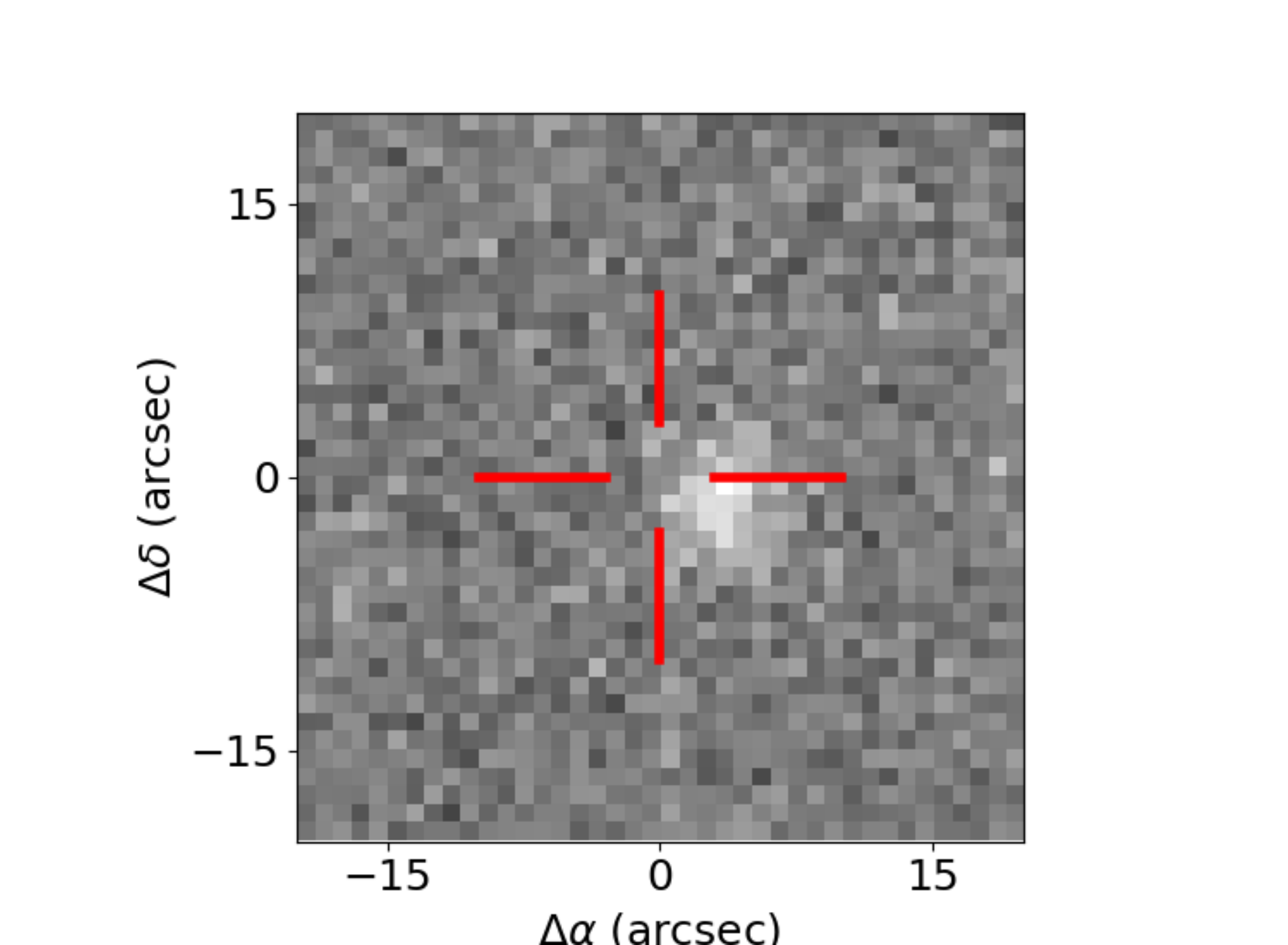}
        \includegraphics[width=0.32\textwidth]
        {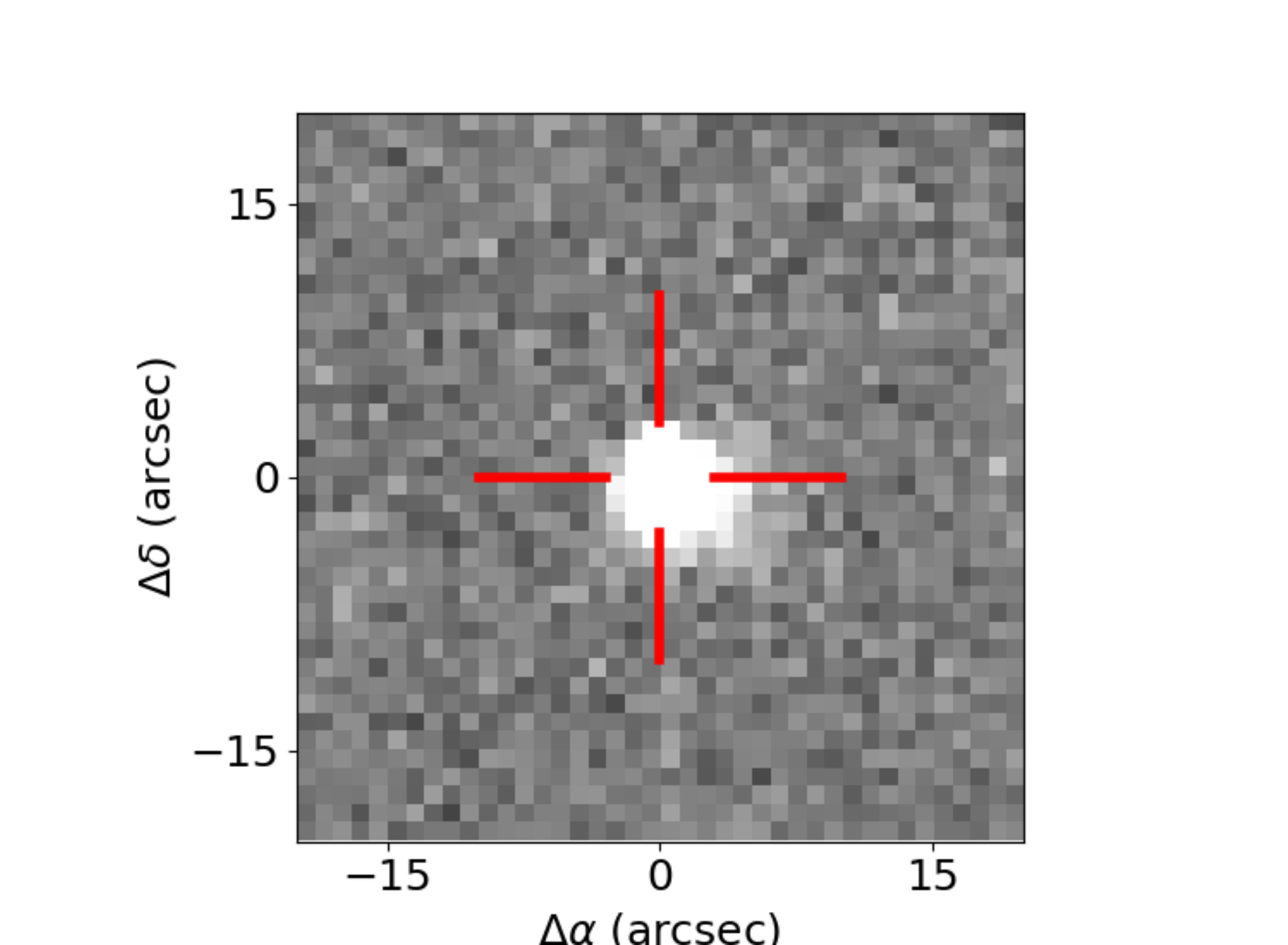}
        \includegraphics[width=0.32\textwidth]
        {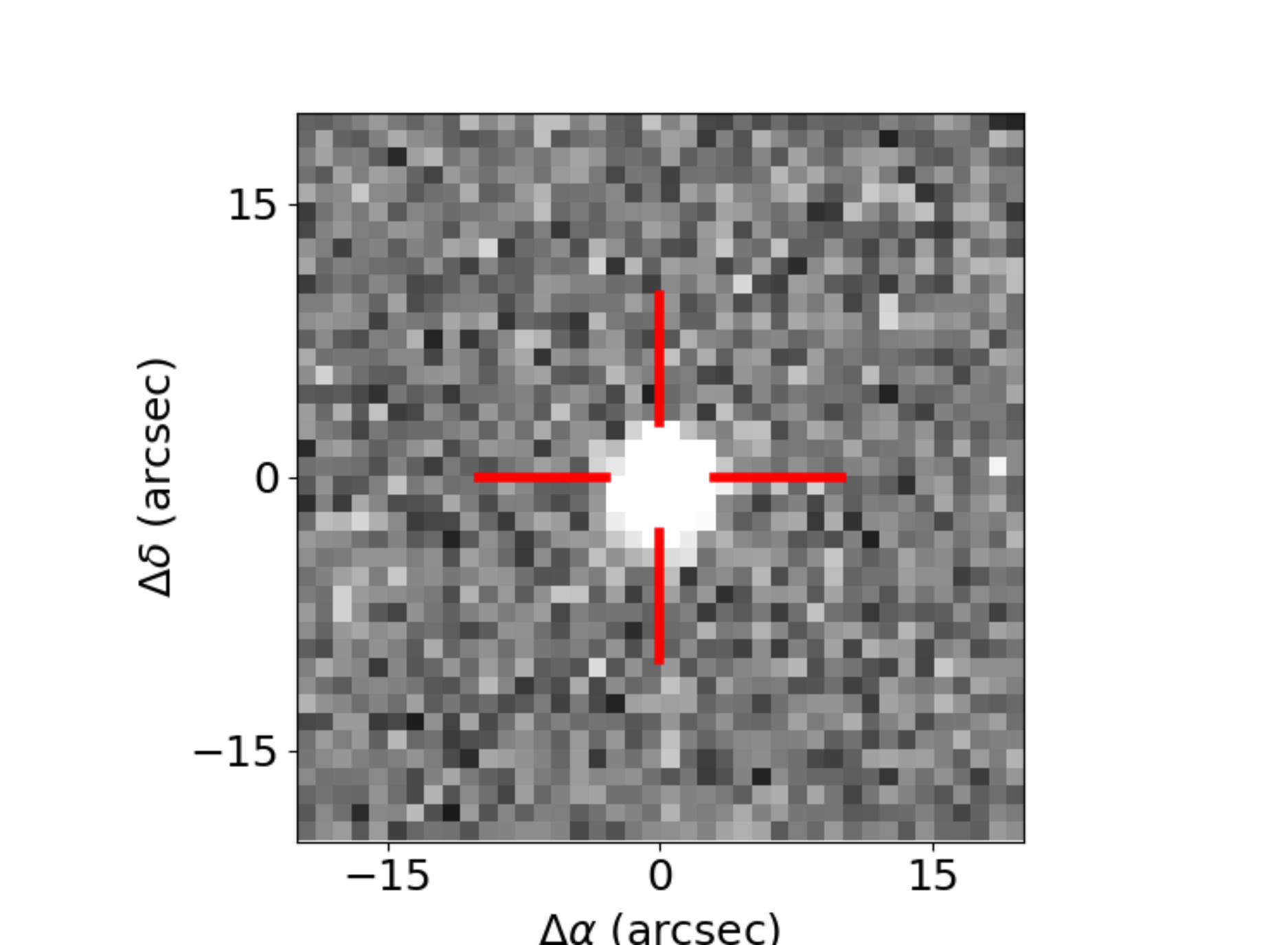}
    } \\
    \subfloat[Dimmer fake transient]{%
        \includegraphics[width=0.32\textwidth]
        {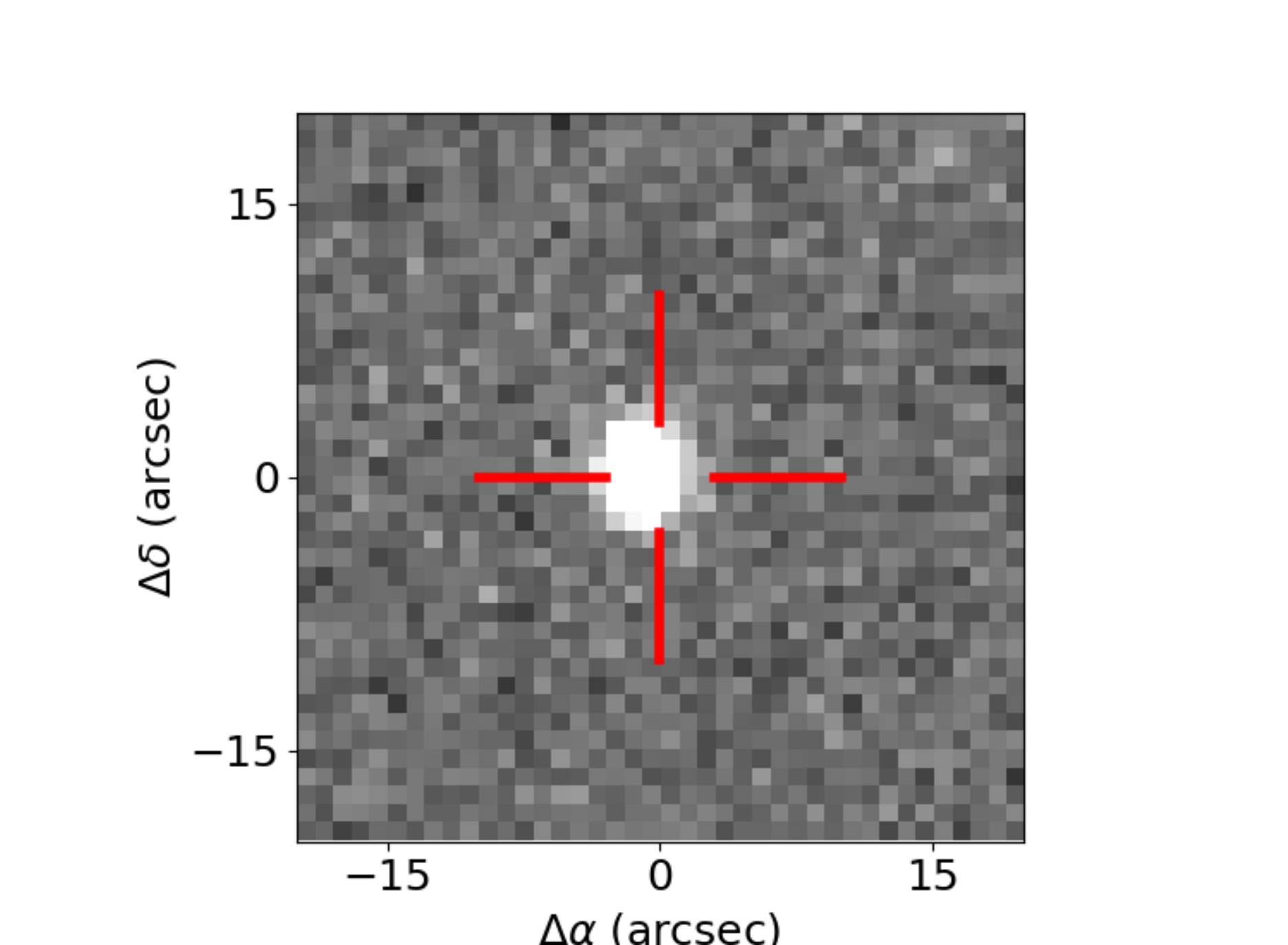}
        \includegraphics[width=0.32\textwidth]
        {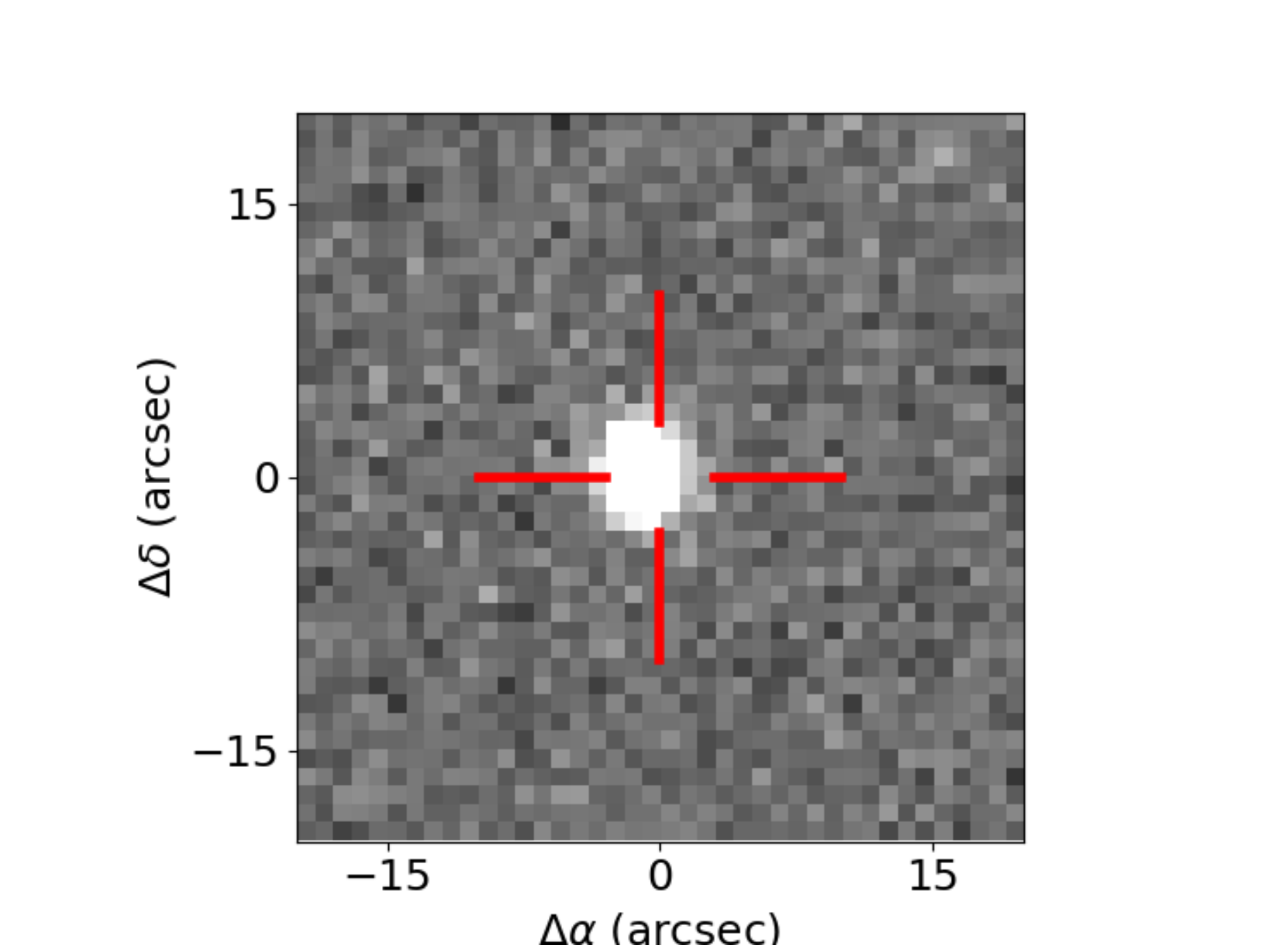}
        \includegraphics[width=0.32\textwidth]
        {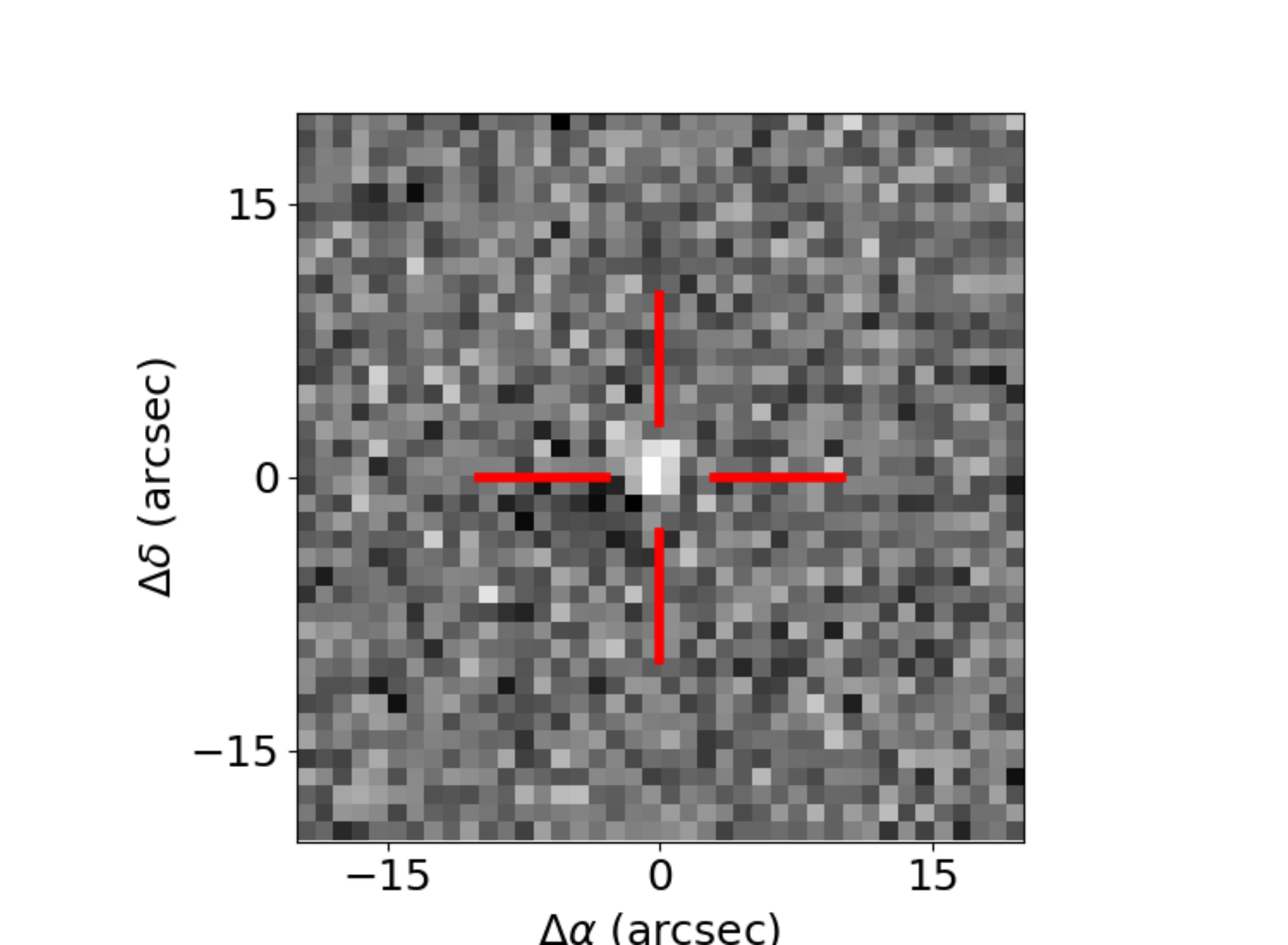}
    }
\end{center}
\caption{An example of an injected transient and the corresponding
difference image thumbnail obtained after the image subtraction.
The leftmost thumbnail (both panels) is from the original image, the middle thumbnail
is a result after a transient is injected, the right thumbnail shows the
difference image. The location of the cross-hair is the approximate point
where the transient was injected.
}
\label{fig:injections_clone_stamping}
\end{figure*}

The fake injections are \emph{bright} stars chosen from each original
image. These are objects having the following
properties:
\begin{align}
\begin{split}
    \mstar \in [13.5, 16] \quad;\quad &\mclst \in [0.5, 1.0] \\
    \mfwhm \in [1.0, 3.0] \quad;\quad &\mellip \in [0.0, 0.3].
\end{split}
\label{eq:injections_bright_stars}
\end{align}
Here $\mstar$ is the apparent magnitude, $\mclst$ is a quantity
having a value between 0 (not star-like) and 1 (star like). $\mfwhm$ is the
full width at half maximum, in pixels. $\mellip$ is the ellipticity of the
object. These quantities are reported after running \texttt{SExtractor} 
\citep{sextractor} on the original images.
\added{
The reason we choose objects in this range is because we want the point
spread function (PSF) to be well estimated, which is the case for bright stars
having a high signal to noise ratio $\gtrsim 100$ ($\mstar \leq 16$). At the same
time we want to avoid pixel saturation and therefore select stars with
$\mstar \geq 13.5$.
}
Objects falling in a 50 pixel
wide edge boundary are left out since they could potentially be affected
by image subtraction artifacts.

A square of side length $\sim 9$ arc seconds
\footnote{More precisely, 9 pixels. $1\text{ pix.} \approx 1.01''$.}
, centered around the star and
local-background subtracted, constitutes a \emph{stamp}. A stamp containing
any other object apart from the source star is avoided. The local-background
refers to that reported by \texttt{SExtractor}. The stamp is scaled by an
appropriate scaling factor to create a point source transient of desired
magnitude. Each transient is allocated a host galaxy
\footnote{About 50 fake transients were injected in each image; $90\%$
having an associated host galaxy, $10\%$ away from any host galaxy.
In this study we only use the injections in host galaxies.}.
We follow \cite{frohmaier_2017} regarding the location in the host
and place our stamp at a random pixel location within a elliptical
radius \footnote{\texttt{KRON\_RADIUS} in \texttt{SExtractor}} of 3 pixels.
This value contains sufficient amount of the flux from the galaxy.

This procedure is performed on all the images in field 100019 of iPTF,
ten-fold, with a total of \replaced{$\approx 2.4 \times 10^6$}{$\approx 2.24 \times 10^6$}
injected transients. The transient magnitudes are chosen uniformly between 15th and
22nd magnitude with the constraint that the stamp is one magnitude fainter
than the original star. 
 \added{
We only re-scale to fainter magnitudes because we do not want artifacts
like noise residuals from the average background subtraction to be scaled
up as noise spikes.
}
Therefore, $\mstamp$ follows:
\begin{equation}
    \mstamp \sim 
    \begin{cases}
    U(15, 22) &; \mstar \in (13.5, 14) \\
    U(\mstar + 1, 22) &; \text{otherwise}
    \end{cases}.
    \label{eq:injections_m_stamp_distribution}
\end{equation}
An example of an injected transient in a galaxy and the new object recovered
by the ISP is shown in Fig.~\ref{fig:injections_clone_stamping}.

\deleted{\texttt{subsubsection}, in favor of \texttt{subsection}.}
\subsection{Recovery Criteria}\label{sec:recovery_criteria}

\begin{figure}[htp]
	\begin{center}
	    \includegraphics[width=1.0\columnwidth, trim=0cm 1cm 0cm 0cm]
	                    {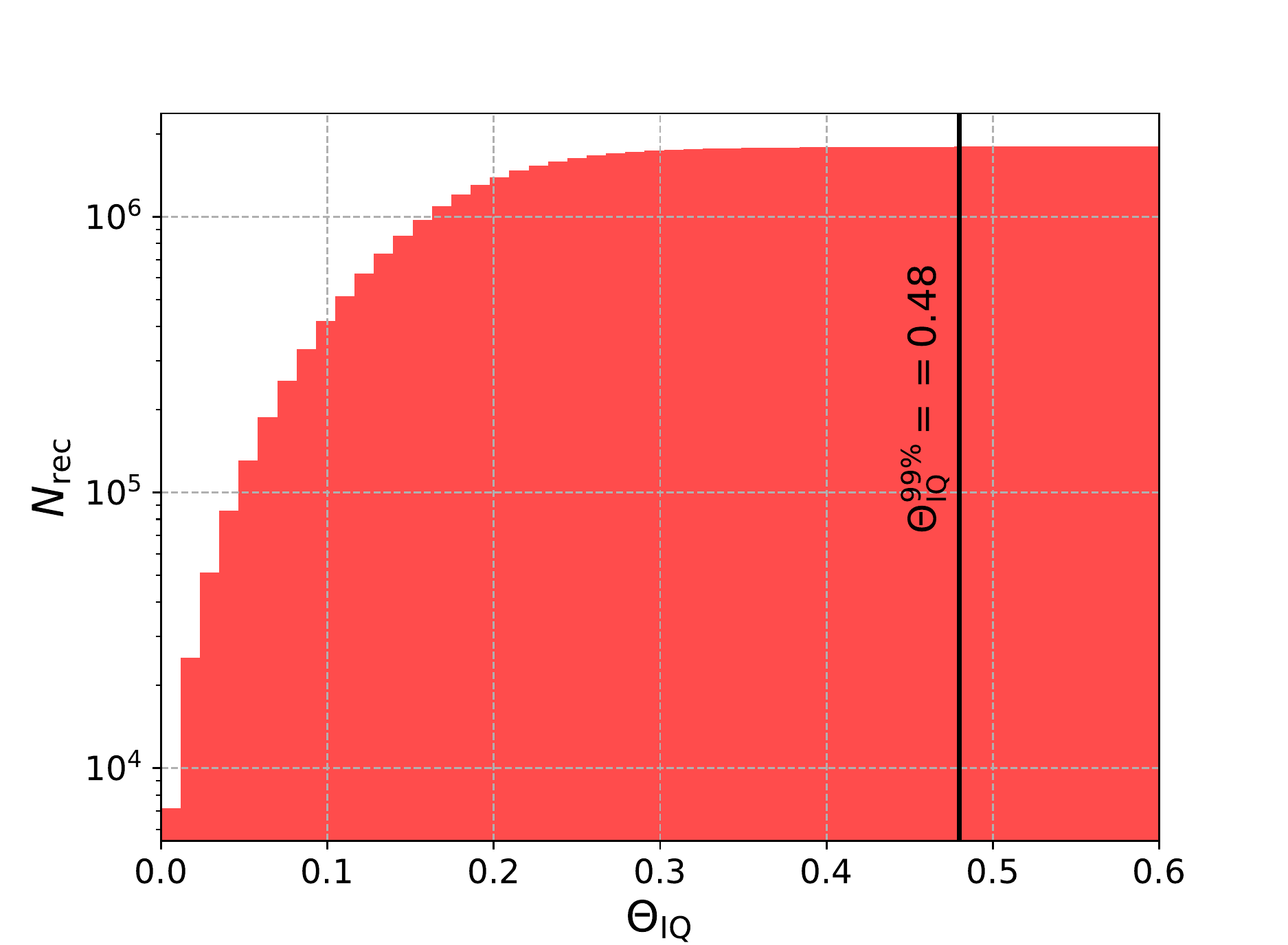}
	\end{center}
	\caption{The figure shows the cumulative histogram of the
	quantity $\thetaiq$, defined as the ratio between the astronomical seeing
	of the image to that of the reference image as given in Eq.(\ref{eq:recovery_theta_iq}).	
	The threshold value $\Theta^{99\%}_{\mathrm{IQ}} = 0.48$ corresponds to the $99\%$
	percentile. We place a constraint of this value when the objects recovered by
	the pipeline are spatially cross matched to an injected transient.
	}
	\label{fig:recovery_theta_iq}
\end{figure}

The recovery efficiency $\varepsilon$ is defined as the
ratio of the number of injections recovered in a part of the parameter space
to the total number of injections in that part. Let our injections be
described by parameters $\blambda$, then:
\begin{equation}
    \varepsilon(\blambda) = \ffrac{\Nrec(\blambda) \mathrm{d}\blambda}
                               {\Ntot(\blambda) \mathrm{d}\blambda}
\label{eq:recovery_efficiency_definition}
\end{equation}
The quantity in the numerator and denominator is the number of
recovered and total injections respectively
$\in \left(\blambda, \blambda + \mathrm{d}\blambda\right)$. Here
$\blambda$ includes both intrinsic source properties of the transient
and its environment along with the observing conditions. Examples of
intrinsic properties include the magnitude of the transient and the surface
brightness of the host galaxy where as those for observing conditions include
airmass or sky brightness. While we control fake transient brightness,
the observing conditions are those of the images themselves.
Since images across the full survey time
are used, the parameter space of the observing conditions is automatically
spanned.

We determine recovery based on the spatial cross matching of the injections
with new objects reported after running the ISP. To determine the tolerance
to be imposed during the cross-matching, we define $\thetaiq$ as: 
\begin{equation}
    \thetaiq = \frac{ \sqrt{(x_{\text{inj}} - x_{\text{rec}})^2 +
                            (y_{\text{inj}} - y_{\text{rec}})^2}
                    }
                    {\Phi}
\label{eq:recovery_theta_iq}
\end{equation}
where $\thetaiq$ is the distance between the injected and the recovered sources
in units of the seeing,  $\Phi$.

We choose the threshold of $\thetaiq$ such that $99\%$ of the found injections
lie within this threshold, which has a value of $\Theta^{99\%}_{\mathrm{IQ}} = 0.48$
(see Fig.~\ref{fig:recovery_theta_iq}). We also impose real-bogus score
threshold, $\rb \geq 0.1$ on the new object. This threshold on $\rb$ is
inspired from survey operation thresholds.
Out of the \replaced{$\approx 2.4 \times 10^6$ injections, we recover $\approx 1.7 \times 10^6$}
{$\approx 2.24 \times 10^6$ injections, we recover $\approx 1.62 \times 10^6$}.

%% file: efficiency.tex
In this section we discuss the results of the injection campaign
mentioned in Sec.~\ref{sec:injections}. We first show some of the
\emph{single parameter} efficiencies as a comparison with those obtained
for PTF (see Fig. 5 of \cite{frohmaier_2017}). For the joint
multi-dimensional detectability, our analysis differs from \cite{frohmaier_2017}.
We treat the problem of detecting a transient in a single epoch as a \emph{binary
classification} problem and use the machinery of supervised learning to
predict whether a transient is detected in that epoch.

\subsection{Single Parameter Efficiencies}
\begin{figure*}[htp]
\begin{center}
    \subfloat{%
        \includegraphics[width=0.44\textwidth , trim=1cm 0cm 0cm 3.5cm]
        {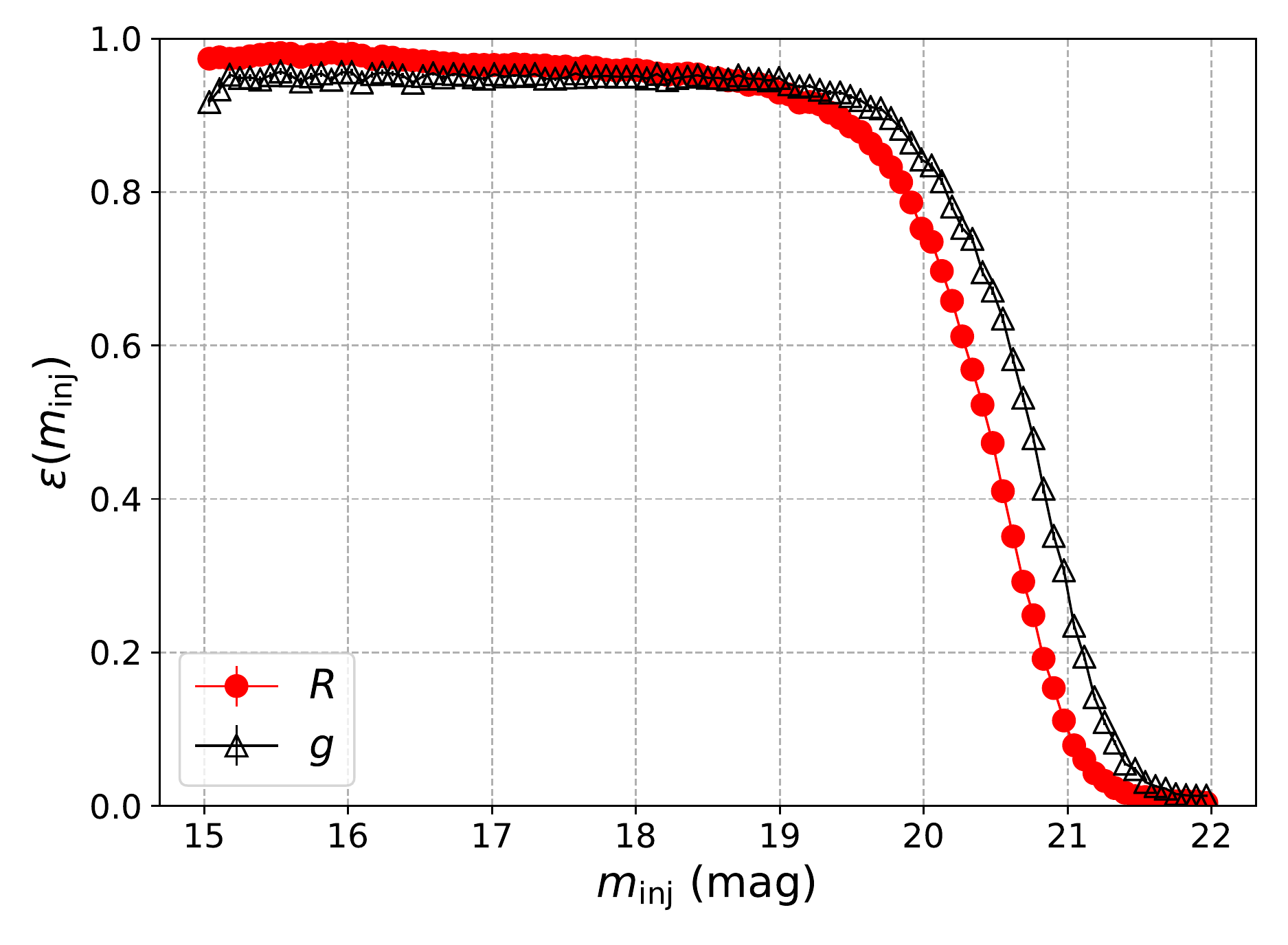}
    }
    \subfloat{%
        \includegraphics[width=0.44\textwidth , trim=0cm 0cm 1cm 3.5cm]
        {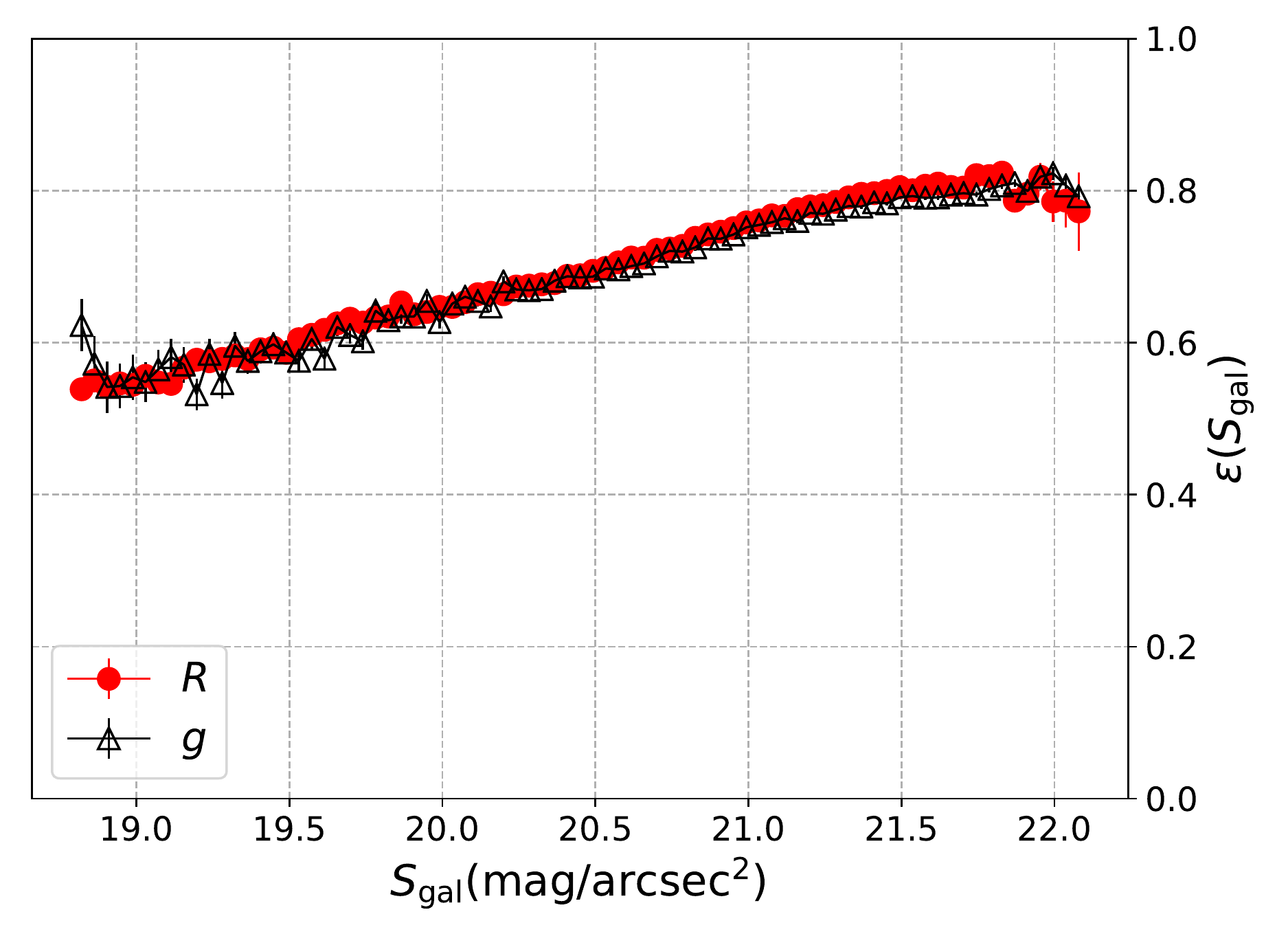}
    } \\
    \subfloat{%
        \includegraphics[width=0.44\textwidth , trim=1cm 0cm 0cm 1cm]
        {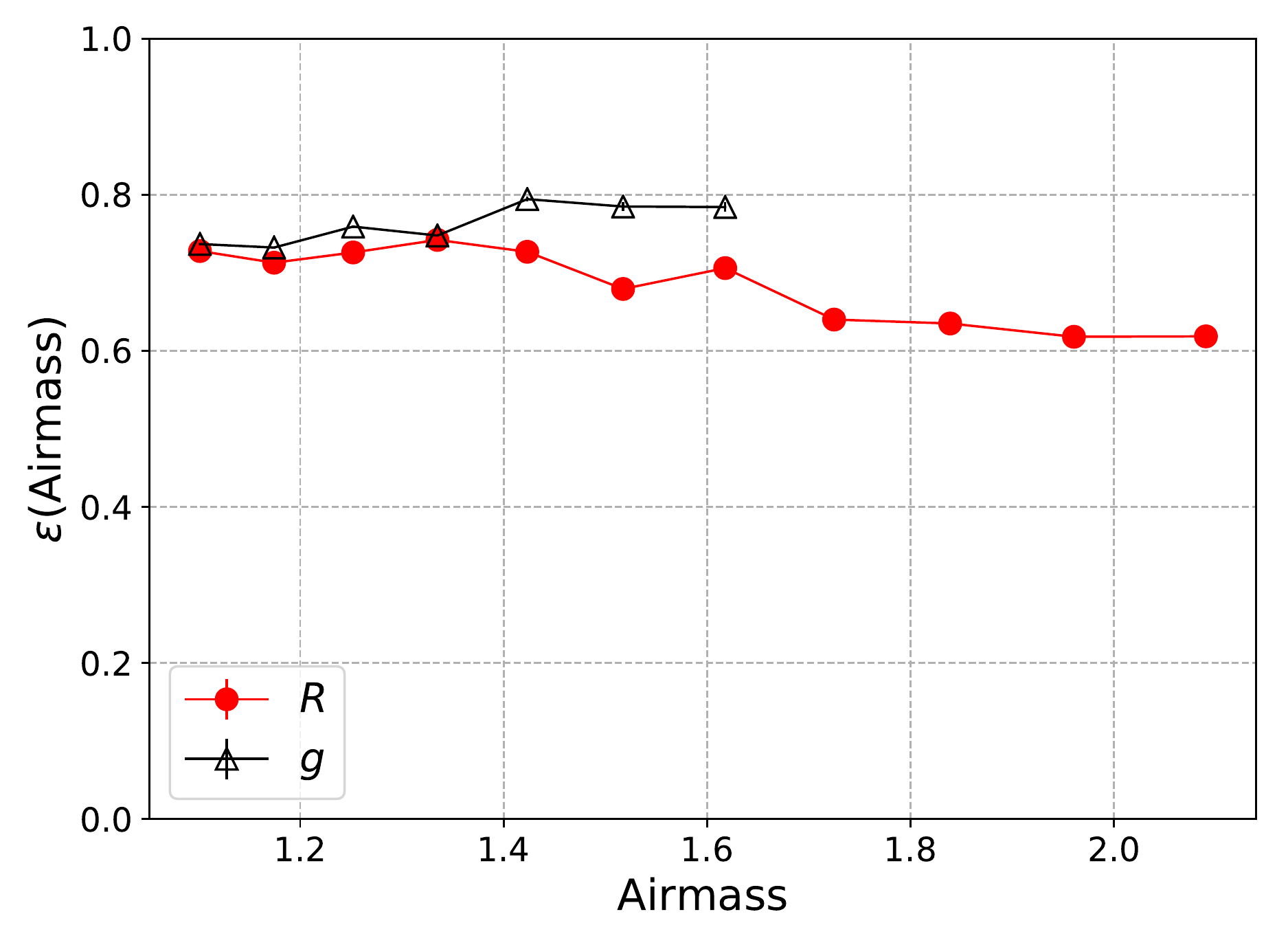}
    }
    \subfloat{%
        \includegraphics[width=0.44\textwidth , trim=0cm 0cm 1cm 1cm]
        {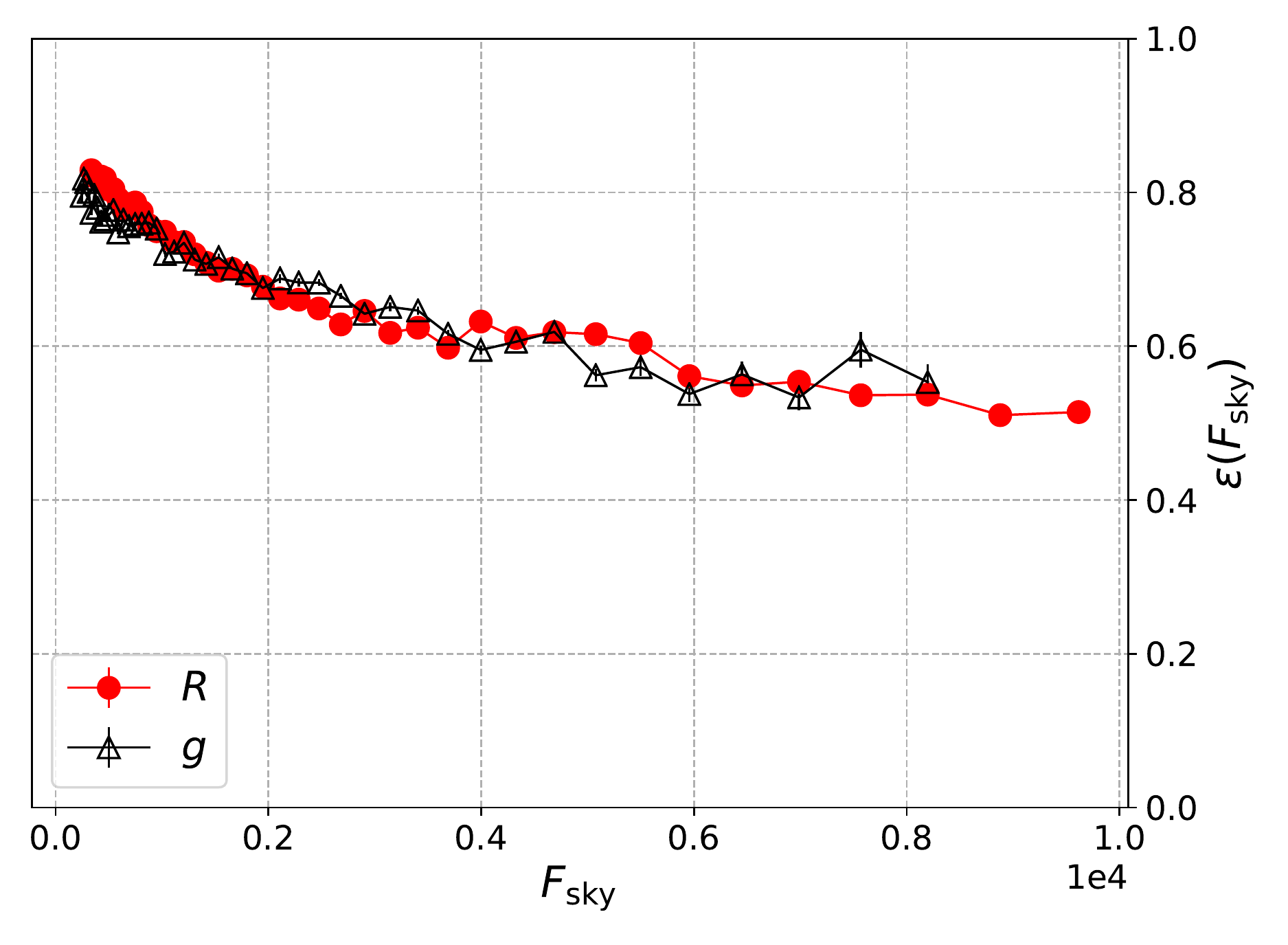}
    } \\
    \subfloat{%
        \includegraphics[width=0.44\textwidth , trim=1cm 0cm 0cm 1cm]
        {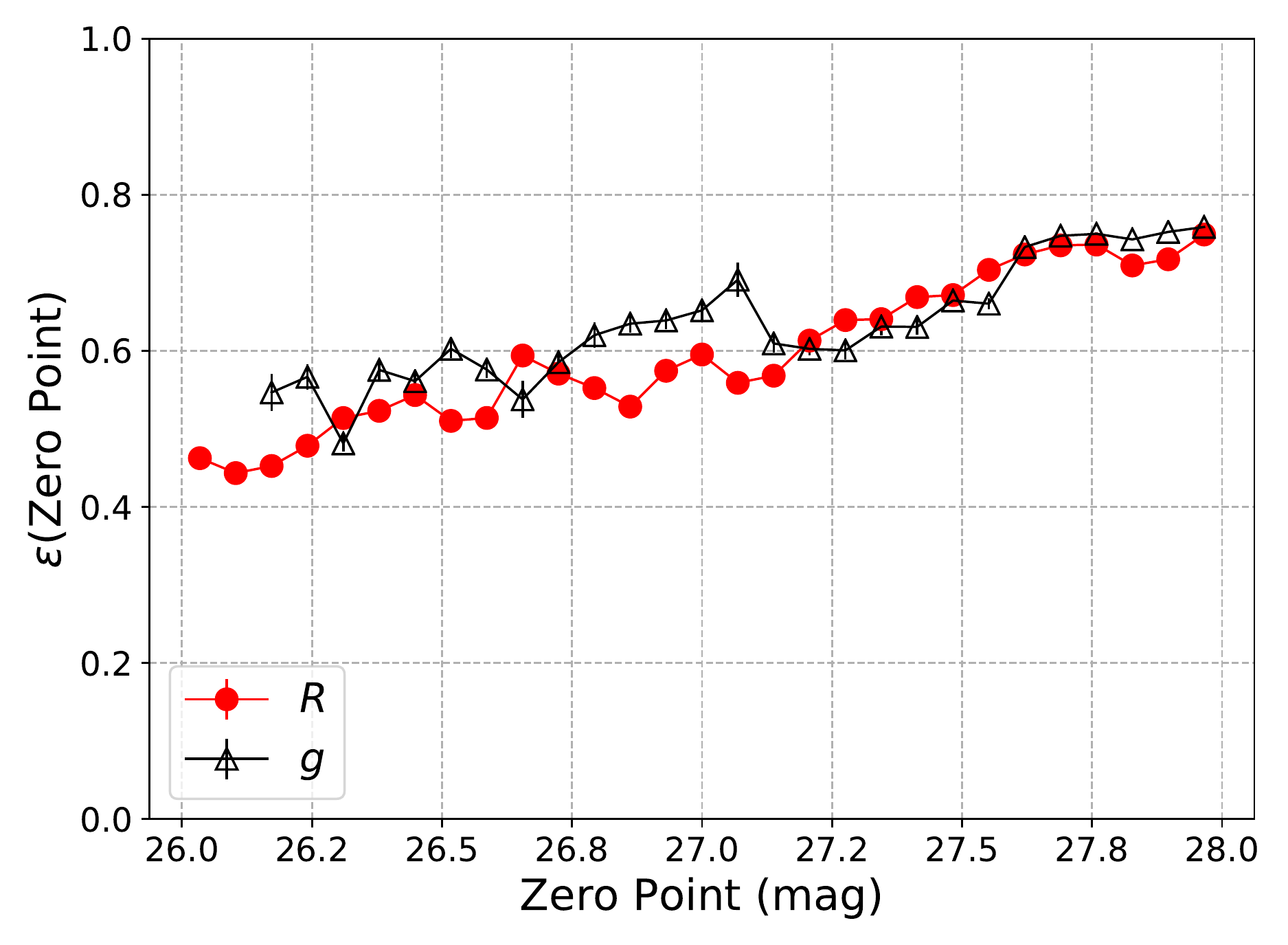}
    }
    \subfloat{%
        \includegraphics[width=0.44\textwidth , trim=0cm 0cm 1cm 1cm]
        {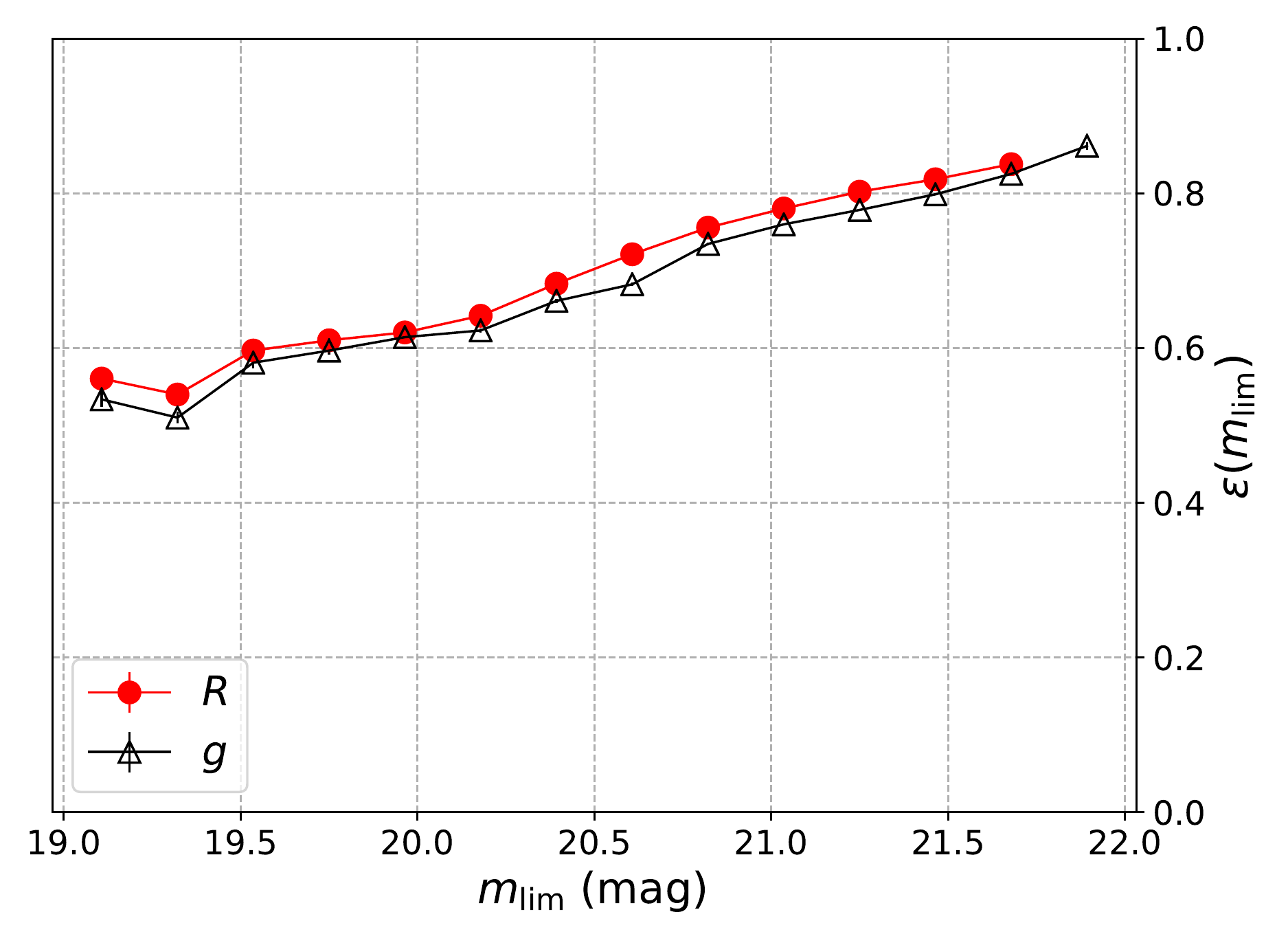}
    } \\
    \subfloat{%
        \includegraphics[width=0.44\textwidth , trim=1cm 1.5cm 0cm 1cm]
        {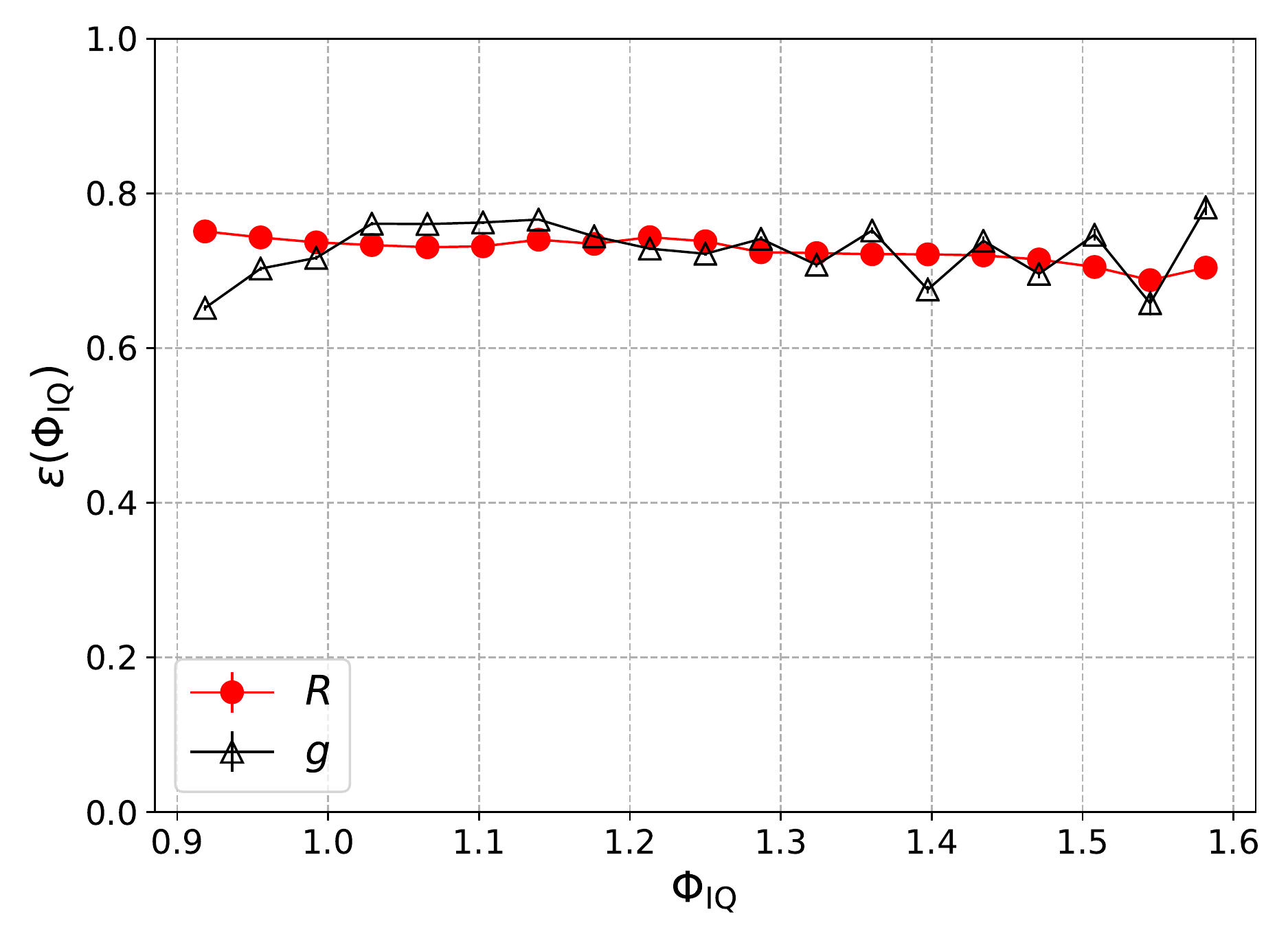}
    }
    \subfloat{%
        \includegraphics[width=0.44\textwidth , trim=0cm 1.5cm 1cm 1cm]
        {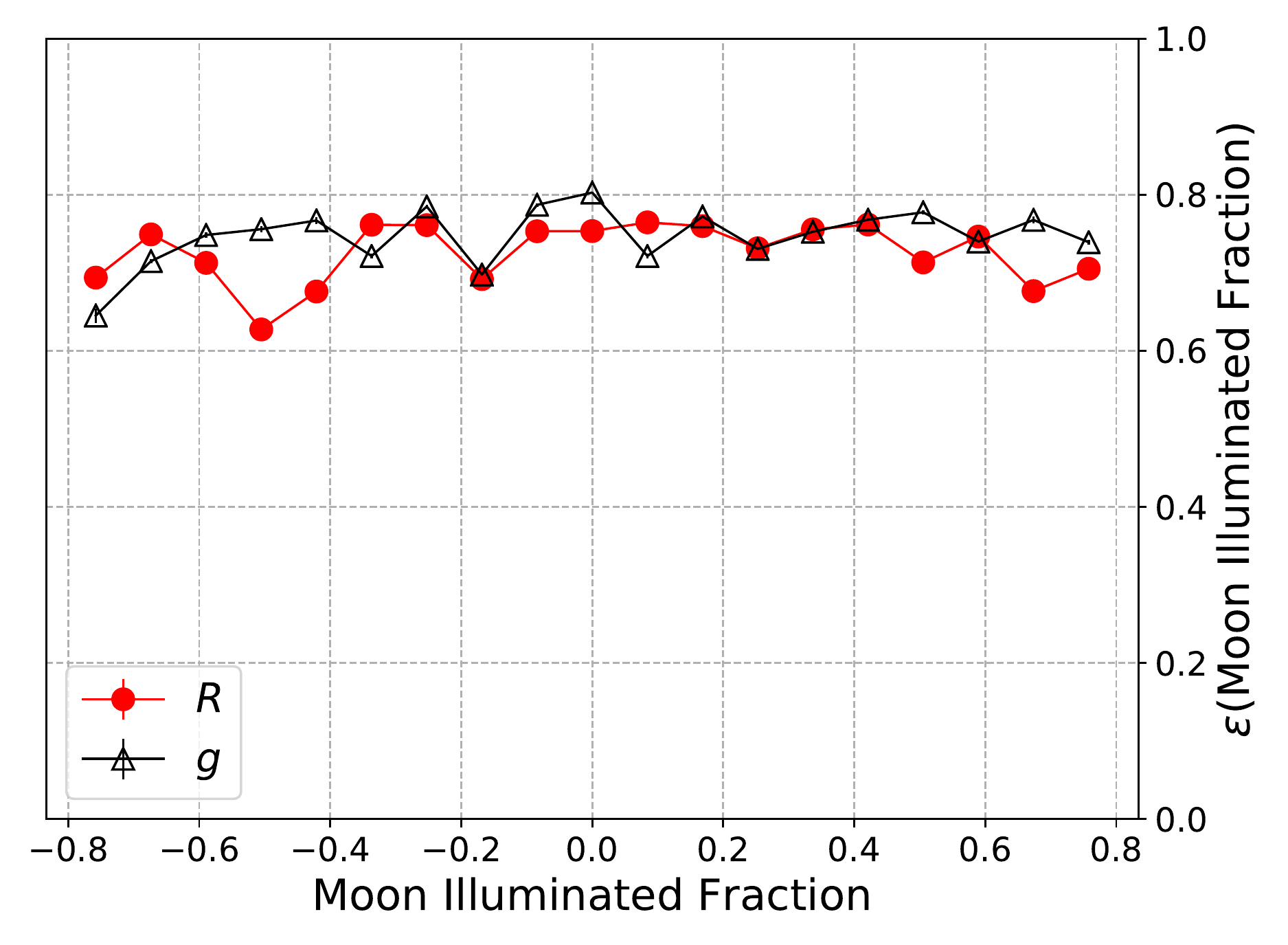}
    }
\end{center}
\caption{The single parameter efficiencies, defined in
Eq.~\ref{eq:efficiency_single_parameter_definition} are shown here. In each of
the panel, the x-axis is the parameter of interest. The top two panels
are parameters which are the intrinsic properties while the remaining
are those taken from observing conditions. We also separate out the
efficiencies based on the filter. While small deviations exists in the
curves the general trend is unchanged based on the filter.
Since there were more number of images (almost 3 times for
field 100019) taken in \emph{R} filter than \emph{g} filter
during iPTF survey, the range of observing conditions
are larger for the \emph{R} filter.
}
\label{fig:efficiency_single_parameter}
\end{figure*}
The \emph{single parameter} efficiency is the marginalized version
of Eq.~(\ref{eq:recovery_efficiency_definition}).
Suppose our parameter of interest is $\theta$ and
the other ``nuisance'' parameters are given by $\bgamma$, such that in
Eq.~(\ref{eq:recovery_efficiency_definition}),
$\blambda = \{\theta, \bgamma \}$. The single parameter efficiency is:
\begin{equation}
    \varepsilon(\theta) = \ffrac{\left[\int_{\bgamma}
                             \Nrec(\theta, \bgamma) \mathrm{d}\bgamma\right]
                                                    \mathrm{d}\theta}
                            {\left[\int_{\bgamma}
                             \Ntot(\theta, \bgamma)\mathrm{d}\bgamma\right]
                                                   \mathrm{d}\theta}
    \label{eq:efficiency_single_parameter_definition} 
\end{equation}

In Fig.~\ref{fig:efficiency_single_parameter} we show the single parameter
efficiencies. The expected trend of missing \added{faint} transients \deleted{at high apparent magnitude}
is seen in the plot for $\mstamp$. We find that the recovery efficiency
starts to drop for transients by the 20th magnitude and sensitivity is almost
nil by the 22nd magnitude.

\subsection{Multi-dimensional Detectability}\label{sec:multi_dim_efficiency}
In this section, we make a selection of parameters from the full parameter set,
$\blambda$, to those on which the detectability depends strongly. In other
words, the detectability is a multi-variate function of all the possible
parameters which influences the detection of a transient. We identify
the minimal set which captures maximum variability. There can be
correlations among a pair of parameters. For example, the sky-brightness,
$F_{\text{sky}}$ and the limiting magnitude, $\mlim$, are correlated - a bright
sky hinders the depth and results in a low value of limiting magnitude.
The variation of the marginalized efficiencies shown in
Fig.~\ref{fig:efficiency_single_parameter} assist us with the choice of such
a parameter set. Since the trend in the single parameter efficiencies are
similar to those from PTF, we select the parameters considered by
\cite{frohmaier_2017} with a minor difference in the usage of the galaxy
surface brightness directly, as used in \cite{Frohmaier_2018}, in place of the
$F_{\mathrm{box}}$
\footnote{Background subtracted flux in a 3x3 box in the location of transient.}
parameter used in the former. This is justified because our fakes were injected in galaxies.

\deleted{subsubsection \textbf{Relevant Parameters}}
\begin{figure}
    \begin{center}
        \includegraphics[width=1.0\columnwidth , trim=1cm 1.5cm 0cm 0cm]
        {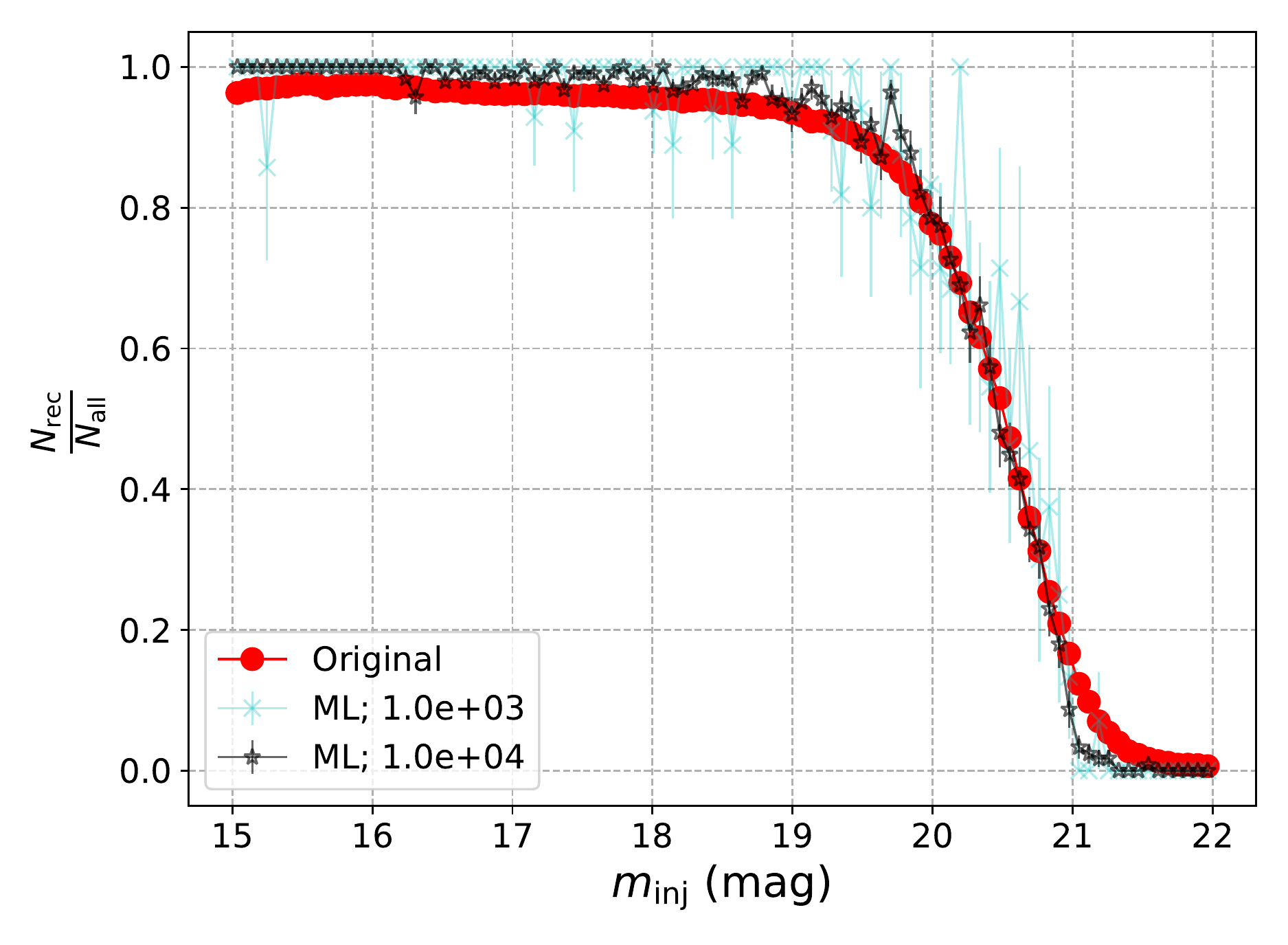}
    \end{center}
\caption{Comparison between single parameter efficiency of transient brightness as
predicted by trained single-epoch classifier in Eq.~(\ref{eq:efficiency_epsilonhat_definition})
versus the distribution obtained from the ISP. The original curve has 
\replaced{
$\approx 2.4\times 10^6$ points, equal to the number of fake simulated transients
}
{$\sim 10^6$ points used to train the classifier}.
The ML curves are made by binning the predictions made by the
single-epoch classifier on a few thousand random points sampled from
the parameter space of the injections (see Eq.~(\ref{eq:efficiency_multidim_parameters})).
Two cases for $10^3$ and $10^4$ points are shown. We see that the
behavior of the classifier converges to that of the ISP within
a small sample size ($\lesssim 1\%$ compared to the size of
original distribution; see Appendix~\ref{sec:appendix} for other parameters)
}
\label{fig:efficiency_ML_performance}
\end{figure}

\begin{table}
    \begin{center}
    \begin{tabular}{ccc}
    \hline
    Training $\%$ & Testing $\%$ & Avg. mis-classification\\
    \hline
    \hline
    75 $\%$       & 25 $\%$      & 5.776 $\%$\\
    80 $\%$       & 20 $\%$      & 5.760 $\%$\\
    85 $\%$       & 15 $\%$      & 5.745 $\%$\\
    90 $\%$       & 10 $\%$      & 5.758 $\%$\\
    \hline
    \end{tabular}
    \end{center}
    \caption{The table shows the average misclassification obtained for
    the \texttt{KNearestNeighbor} classifier. The complete dataset contains
    $\approx 2.24 \times 10^6$ fake point source injections of which
    $\approx 1.62 \times 10^6$ ($\approx 6.2 \times 10^5$) are found (missed)
    by the ISP. This is split into respective training and testing
    fractions. The right-most column shows the fraction of the testing set
    for which the predictions made by the classifier, trained on the corresponding
    training fraction differed from the actual value.
    The misclassification does not change significantly as the size of training
    data is varied and is attributed mostly to systematics. We quote a conservative
    value of $6\%$ as the systematic uncertainty of the classifier. 
    }
    \label{tab:efficiency_avg_misclassification}
\end{table}

We choose, the following set to represent the dependence of detectability:
\begin{equation}
    \bbeta = \{m, \shost, \fsky, \phiiq, \mlim\}.
    \label{eq:efficiency_multidim_parameters}
\end{equation}
Here $m$ is the apparent magnitude of the transient, $\shost$ is the host galaxy
surface brightness, $\fsky$ is the sky brightness, $\phiiq$ is the ratio of the
astronomical seeing to that of the reference image and $\mlim$ is the limiting
magnitude. The quantities $m$ and $\shost$ are natural in capturing detectability.
Sky brightness affects the detectability in a strong
way, as is apparent from Fig.~\ref{fig:efficiency_single_parameter}.
The $\phiiq$ parameter captures the variability of the atmosphere.
Finally, the limiting magnitude, $\mlim$, although correlated with $\fsky$,
captures longer exposure times and status of instrument electronics.

With this set, we use the machinery of supervised learning provided by
the \texttt{scikit-learn} library \citep{scikit-learn} to train a binary classifier
based on the results of the ISP. Once trained, the classifier outputs a probability
of detection given arbitrary but physical values of $\bbeta$. We denote this trained
classifier by $\epsilonhat$:
\begin{equation}
    \epsilonhat = \epsilonhat(m, \shost, \fsky, \phiiq, \mlim).
    \label{eq:efficiency_epsilonhat_definition}
\end{equation}
The \texttt{scikit-learn} library provides a suite of
classifiers. We choose the non-parametric \texttt{KNearestNeighbor} classifier
based on speed and accuracy given our large volume of training data.
\added{
Our complete dataset comprises of $\sim 2.24 \times 10^6$ fake point source
injections of which $\sim 1.62 \times 10^6$ ($\sim 6.2 \times 10^5$) are found
(missed) by the ISP.
}
We train the \replaced{model}{classifier} using 11 neighbors - twice the number
of dimensions plus one to break ties. The observation of a fiducial transient is
a point in this parameter space. To decide if that point is ``missed'' or ``found'',
we use a majority vote from the nearest 11 neighbors.
\added{
To cross-validate the performance, the dataset is split into a training
set containing $90\%$ of the full dataset, and a testing set containing the
remaining $10\%$.
}
We checked that increasing the number of neighbors
does not significantly increase the correctness of predictions made by the
classifier. We note that one could use a different threshold for this classification.
For example, a different option could be to use greater than 3 ``found''
neighbors to call the arbitrary point as found. However, it comes at a
cost of misclassification.
\replaced{
We calculate the systematic uncertainty of
the classifier by splitting the original data into training and testing sets.
We find that by using the majority vote criterion the misclassification is
}
{
From the predictions of the classifier on the testing set, we find the
systematic uncertainty of the classifier to be
}
$\approx 6\%$ i.e. 6 out of 100 predictions made by the classifier
is expected to be either true negative or false positive cases.
The result does not change much if the size of the training and testing set is
varied \added{(see Table~\ref{tab:efficiency_avg_misclassification})}.
A comparison between the predictions made by the trained classifier and the
original ISP efficiency with the transient magnitude is presented in
Fig.~\ref{fig:efficiency_ML_performance}.
We see that the behavior of the ISP is reproduced by feeding the
classifier with only a few thousand points randomly chosen from the parameter
space.

%% file: rates.tex
\begin{figure}[htp]
    \begin{center}
        \includegraphics[width=1.0\columnwidth, trim=0.5cm 0cm 0.5cm 0cm]
            {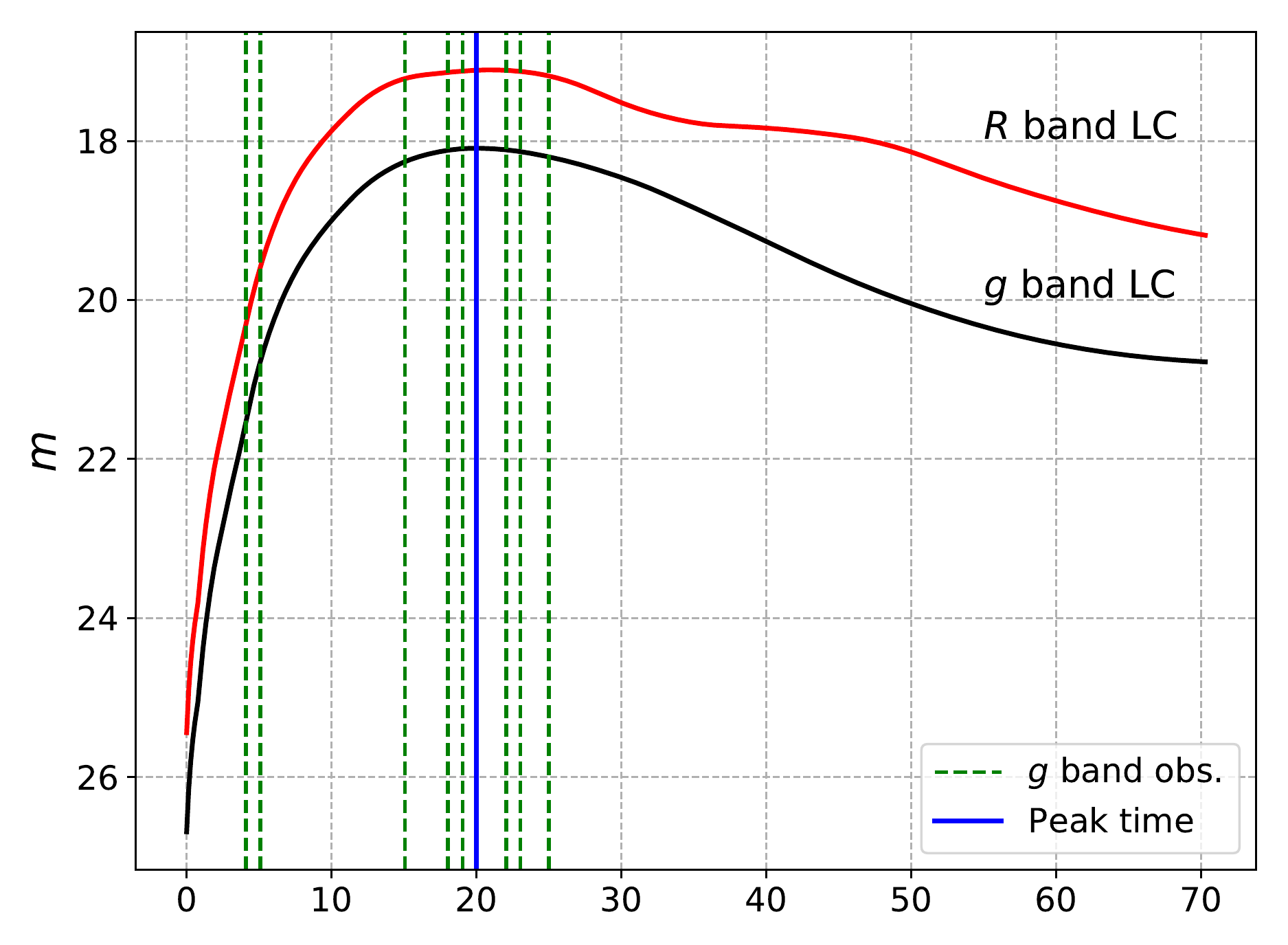}
        \includegraphics[width=1.0\columnwidth, trim=0.5cm 1cm 0.5cm 0cm]
            {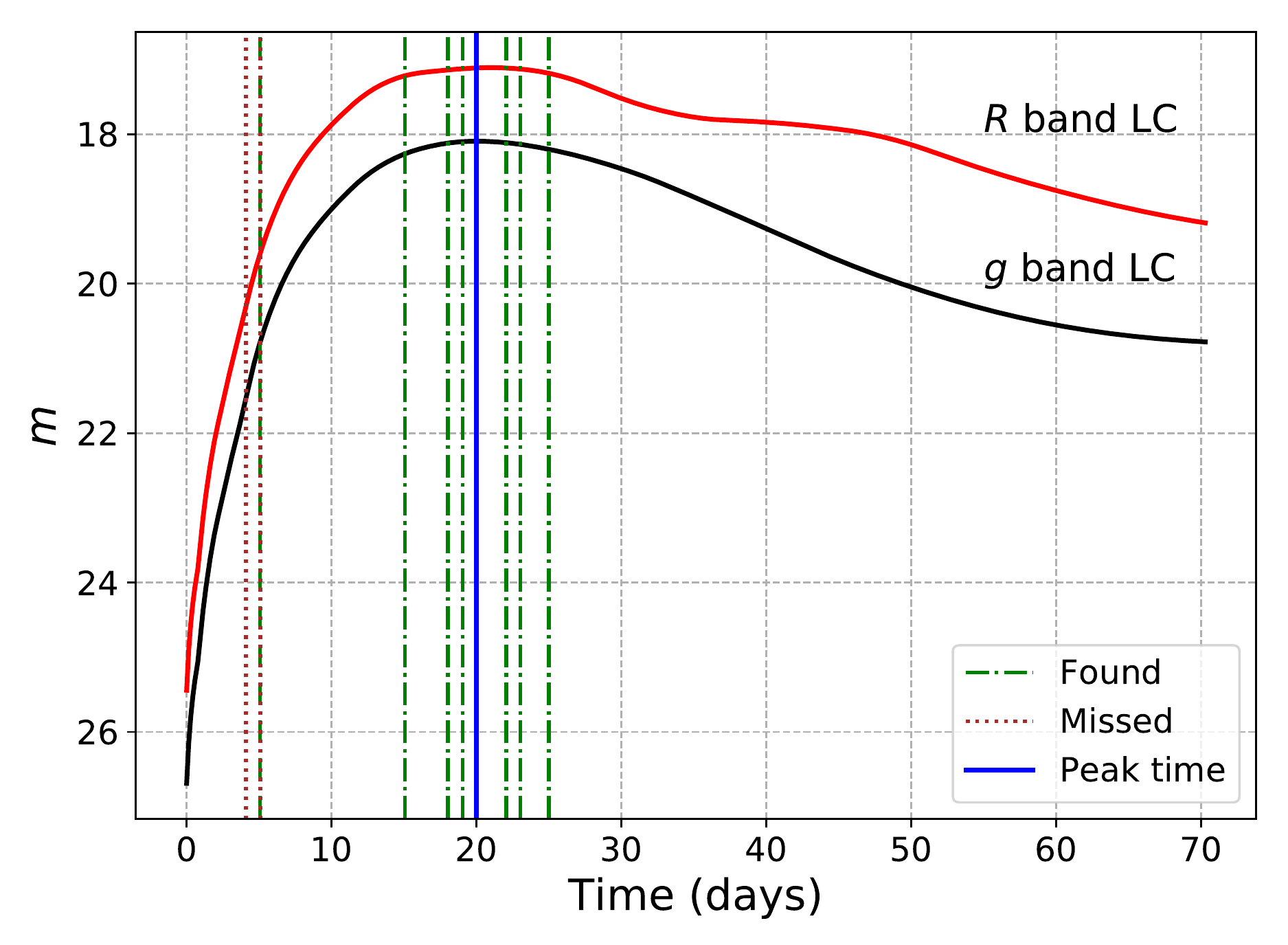}
    \end{center}
    \caption{\textbf{Upper panel}: An example of a SALT2 lightcurve,
    with the apparent magnitude, $m$ on the y-axis and time on the x-axis.
    The lightcurves in the iPTF $R$ and $g$ bands are shown. The
    observations of the telescope are shown as vertical lines. At
    each observation, we also have the observing conditions of the
    telescope from archival data.
    \textbf{Lower panel}: The same lightcurve is plotted, however,
    the vertical lines now represent the detectability from the single epoch classifier.
    Based on the criteria of confirming a lightcurve as SN~Ia, this lightcurve
    was recovered.}
    \label{fig:rates_example_lightcurve}
\end{figure}

\begin{figure}[tph]
	\begin{center}
	    \includegraphics[width=1.0\columnwidth, trim=0.5cm 0.5cm 0.5cm 0cm]
            {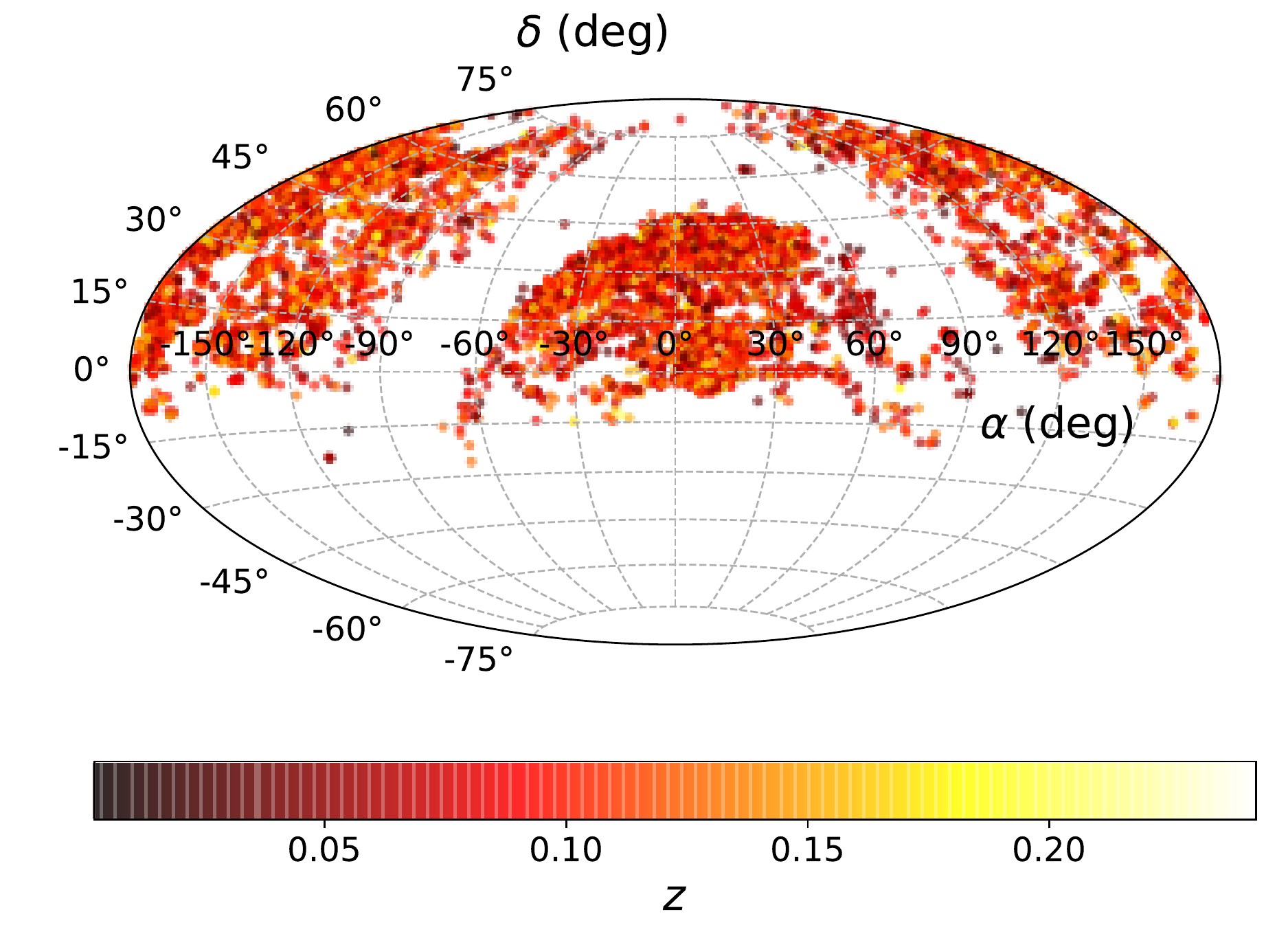}
	\end{center}
    \caption{An ensemble of SN~Ia lightcurves were simulated out to a redshift,
    $z_{\mathrm{max}}^{\mathrm{Ia}} = 0.28$, uniform in co-moving volume. This figure shows
    the distribution of the recovered SN~Ia in the sky colored by the redshift.
    The galactic plane can be seen as the half annulus region with no detections.}
    \label{fig:rates_SNIa_skymap}
\end{figure}

\begin{figure}[tp]
    \begin{center}
        \includegraphics[width=1.0\columnwidth, trim=1.5cm 1cm 0cm 0cm]
            {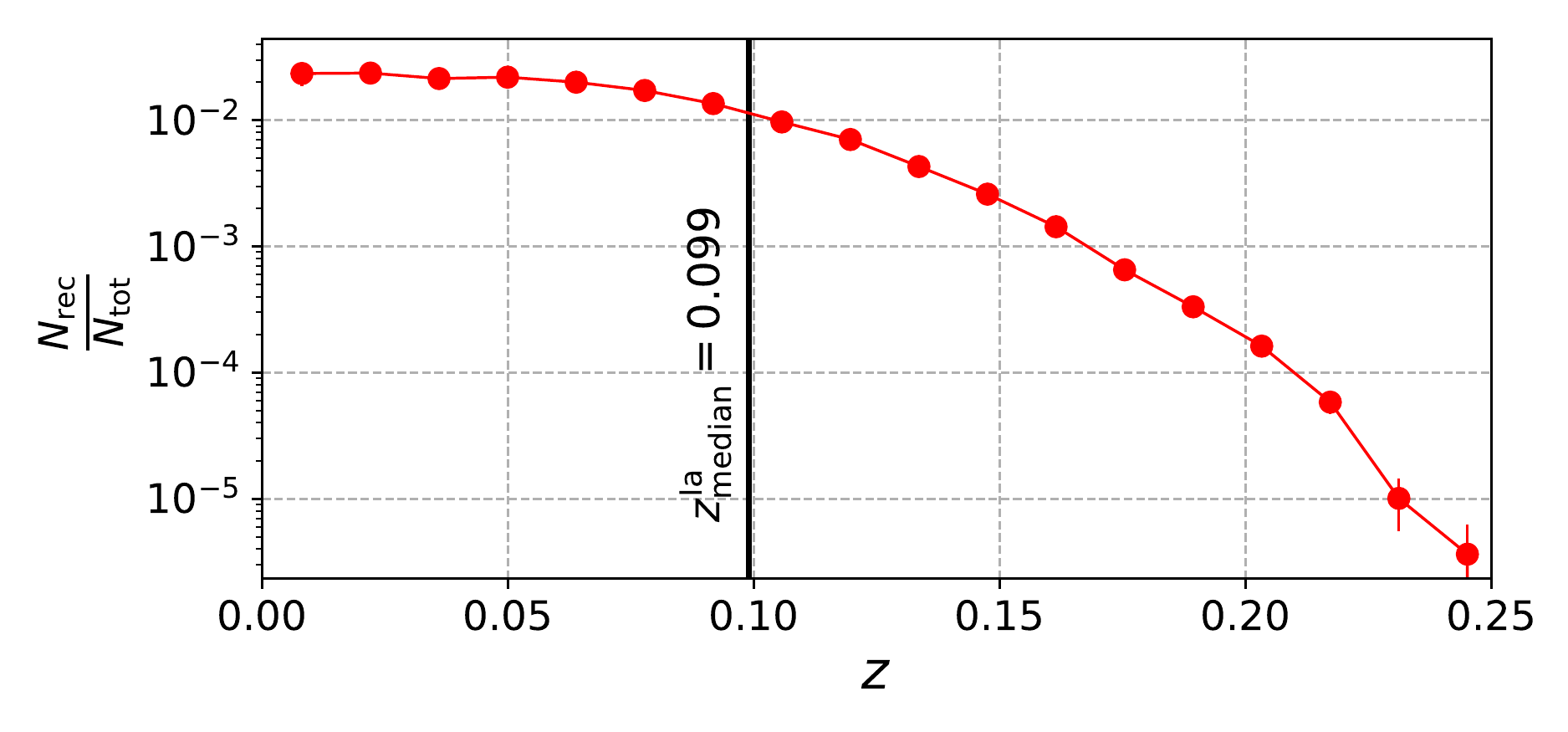}
    \end{center}
    \caption{Recovery efficiency of the SN~Ia lightcurves
    as a function of redshift, $z$. The median volume weighted redshift
    is found to be $z_{\text{median}}^{\text{Ia}} = 0.099$.}
    \label{fig:rates_SNIa_recovery}
\end{figure}
In this section, we assess the detectability of lightcurves using SNe~Ia as
our case study. We simulate lightcurves with varying intrinsic properties,
sky location and redshift, and use the single epoch detectability classifier
mentioned in Eq.~(\ref{eq:efficiency_epsilonhat_definition}) together with the
observing schedule of iPTF to determine their sensitivity.
The steps are as follows:

\begin{enumerate}
    \item We simulate lightcurves of varying intrinsic properties over
          space-time.
    \item From the complete iPTF observing schedule, we determine the
          observations of the evolving lightcurve. This depends on
          the duty cycle of the instrument. On extended periods
          with no observations, the simulated lightcurves are missed.
    \item We associate a host galaxy with the supernova by choosing a
          surface brightness value from the distribution of galaxy
          surface brightness in the survey.
    \item Every time the transient is ``seen'' by iPTF , we feed the
          combination of the apparent magnitude, host galaxy surface
          brightness along with the observing conditions at that epoch
          to the trained single epoch classifier developed in
          Sec. ~\ref{sec:efficiency}. This step, in a sense, mimics the
          action of the ISP.
    \item We call the lightcurve \emph{recovered} when we have at least
          5 found observations, all brighter than 20th magnitude,
          with a minimum of 2 observations on the lightcurve rise and a
          minimum of 2 on the fall. This is motivated by
          survey time discoveries.
\end{enumerate}
We also consider type II supernova lightcurves for comparison.
Type II supernovae are complex and are further categorized into different subtypes.
We consider the IIp subtype because compared to the $\sim$ weeks long variability
of SNe~Ia, IIp lightcurves vary $\sim 100$ days and hence is a complimentary case
to study. The analysis for the IIps, however, is simpler compared to Ias. 

\subsection{SN~Ia Lightcurves}\label{subsec:salt2_lightcurves}
We use SN~Ia lightcurves from the SALT2 model \citep{salt2}.
In particular, we use the \texttt{Python} implementation of SALT2 provided
in \texttt{sncosmo} library \citep{sncosmo}.
This model is based on observations of SNe~Ia by the SDSS and SNLS surveys.
The free parameters of the model include the stretch ($x_1$) and color ($C$)
parameters of the SN~Ia. Regarding the range of these parameters, we follow
same range as \cite{frohmaier_2017} (see Table 1 and Eq.(4) therein).
The ranges cover the possible lightcurve morphologies of SNe~Ia \citep{jla}.
We show an example lightcurve, at a redshift of
$z = 0.01$ with an instrinsic $M_B = -19.05$ in Fig.~\ref{fig:rates_example_lightcurve}.
When propagating the flux, we also take into account the extinction due to
host galaxy dust and the Milky Way (MW) dust. We use the MW dust map
by \cite{f99dust} which is a part of the \texttt{sncosmo} package.
For the host galaxy extinction, we use the distribution of $E(B-V)$ of SN~Ia
in their host galaxies \citep{hatano}. Dust extinction plays a significant
role in the detectability of lightcurves as the SNe can be dimmed by as much
as $1 - 1.5$ magnitudes.

\subsection{Lightcurve Ensemble}\label{subsec:lightcurve_ensemble}
We simulate $\approx 5 \times 10^6$ SN~Ia lightcurves uniformly in co-moving
volume up to redshift, $z_{\mathrm{max}}^{\mathrm{Ia}} = 0.28$
\footnote{
The $z_{\mathrm{max}}^{\mathrm{Ia}} = 0.28$ is high enough to capture the
spacetime boundary of iPTF sensitivity. Also, no simulations are done
below a declination, $\delta_{\mathrm{min}}\approx -31^{\circ}$ consistent
with hardware limitations for iPTF.
},
uniform in peak time distribution in the observer frame.
We assume a flat $\Lambda\mathrm{CDM}$ cosmology with
Hubble constant, $H_0 = 69.3$ $\mathrm{km s^{-1}/Mpc}$ and matter to critical
density, $\Omega_\mathrm{m} = 0.287$ \added{\citep{2013ApJS..208...19H}}
\footnote{\texttt{astropy.cosmology.WMAP9}}.
We associate a host galaxy surface brightness to each of these SNe using the
distribution of surface brightness from iPTF data.

The epochs when the SN~Ia is observed come from the iPTF observing
schedule. At each observation, we obtain the
transient magnitude at that epoch from the lightcurve and the observing
conditions from the iPTF survey database. The single epoch classifier then
tells us the epochs when the transient was detected. An example is shown in
Fig.~\ref{fig:rates_example_lightcurve} where the vertical lines in the upper
and lower panel respectively represent the observations and detections at each
epoch.

\subsection{SN~Ia Space-time Sensitive Volume}
To understand rates, one must have a good estimate of the survey sensitivity to
particular transient types. Let $\lambdasne$ be the expected count of SNe
seen during survey time. Then, with $R$ as the intrinsic
rate we have:
\begin{eqnarray}
    \lambdasne &=& \int f(t;\underbrace{M_B, z, \dots}_{\bkappa})
                \overbrace{\frac{\mathrm{d}N}{\mathrm{d}t_e \mathrm{d}V_c}}^{R}
                \frac{1}{1 + z}
                \frac{\mathrm{d}V_c}{\mathrm{d}z} \mathrm{d}z \mathrm{d}t \mathrm{d}\bkappa
                \nonumber \\
               &=& R \int f(t;\underbrace{M_B, z, \dots}_{\bkappa}) \frac{1}{1 + z}
               \frac{\mathrm{d}V_c}{\mathrm{d}z} \mathrm{d}z \mathrm{d}t \mathrm{d}\bkappa
               \label{eq:rates_Nobs_VT_relation}\\
               &=& R \langle VT \rangle, \nonumber
\end{eqnarray}
where the integral runs over time of observation and co-moving volume up to
$z_{\mathrm{max}}^{\mathrm{Ia}}=0.28$. The \emph{selection} function,
$f(\dots) \in \{0, 1\}$, is to be interpreted as the weight assigned to regions in
space-time. The value of the selection function is a consequence of running a
particular instance of SN~Ia through the observing schedule and inferring detectability
based on the single-epoch classifier in Eq.~(\ref{eq:efficiency_epsilonhat_definition}).
Therefore, the selection function depends on the observer time, $t$, which
captures the duty cycle and cadence. Also, it depends on the intrinsic properties
of the supernova like the absolute intrinsic magnitude, $M_B$, the redshift,
$z$, at which it was simulated, the sky location and so on. These are
collectively represented by $\bkappa$ in Eq.~(\ref{eq:rates_Nobs_VT_relation}).
Since we have distributed the supernovae uniformly in co-moving volume, the
integral is approximated in the Monte-Carlo sense:
\begin{eqnarray}
    \langle VT \rangle &=&
            \int f(t;\underbrace{M_B, z, \dots}_{\bkappa}) \frac{1}{1 + z}
            \frac{\mathrm{d}V_c}{\mathrm{d}z} \mathrm{d}z \mathrm{d}t \mathrm{d}\bkappa \nonumber \\
                       &\approx &
            \frac{\Nrec}{\Ntot}
            T \int \frac{1}{1 + z} \frac{\mathrm{d}V_c}{\mathrm{d}z}
            \mathrm{d}z,
\end{eqnarray}
where $\Nrec$ is the number of SNe recovered from this simulation campaign,
$\Ntot$ is the total number simulated and $T$ is the four year period of iPTF
over which we performed the simulations \footnote{More specifically,
Oct 23, 2012 to Mar 3, 2017 $\Rightarrow$ 1592 days}.
We obtain the result:
\begin{equation}
    \langle VT \rangle_{\mathrm{Ia}} = 2.93 \pm 0.21 \times 10^{-2}
    \; \mathrm{Gpc^{3}\,yr}
\label{eq:rates_VT_value}
\end{equation}
where the error includes the $\sim 1/\sqrt{N}$ statistical error from Monte Carlo
integration and the $6\%$ systematic error of the single epoch detectability classifier
computed in Sec.~\ref{sec:multi_dim_efficiency},
the latter being the dominant source of error.
The distribution of the detected SNe~Ia in sky is shown in Fig.~\ref{fig:rates_SNIa_skymap}
colored by redshift. Using the recovered SNe~Ia, the median sensitive co-moving volume
is found to be $0.305$ $\mathrm{Gpc^3}$. We report the redshift corresponding to
this value as the median sensitive redshift to SNe~Ia,
$z_{\text{median}}^{\text{Ia}} = 0.099$, shown in Fig.~\ref{fig:rates_SNIa_recovery}.

\subsection{SN~IIp Space-time Sensitive Volume}
\begin{figure}[htp]
    \begin{center}
        \includegraphics[width=1.0\columnwidth, trim=0.5cm 0cm 0.5cm 0cm]
            {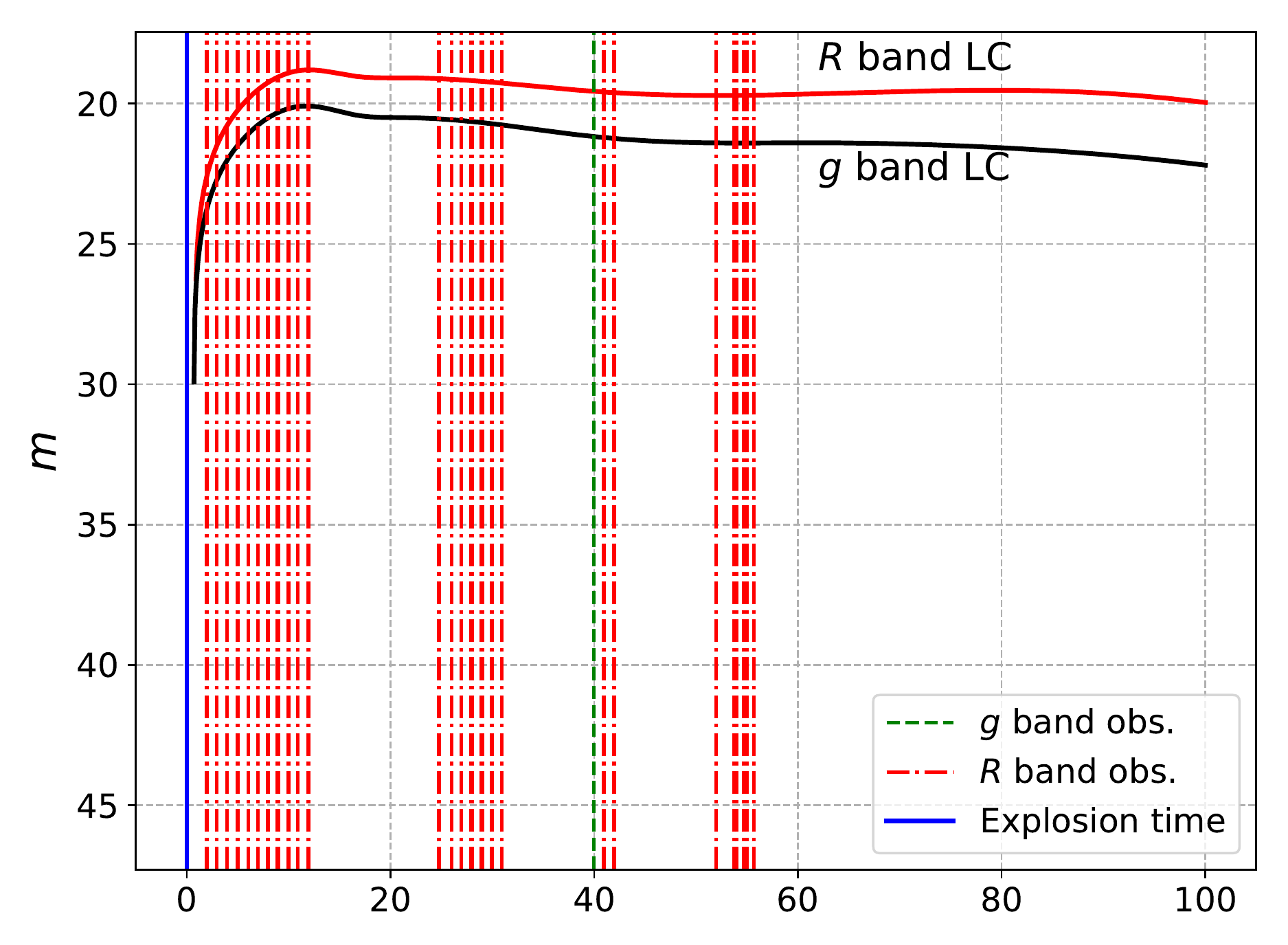}
        \includegraphics[width=1.0\columnwidth, trim=0.5cm 1cm 0.5cm 0cm]
            {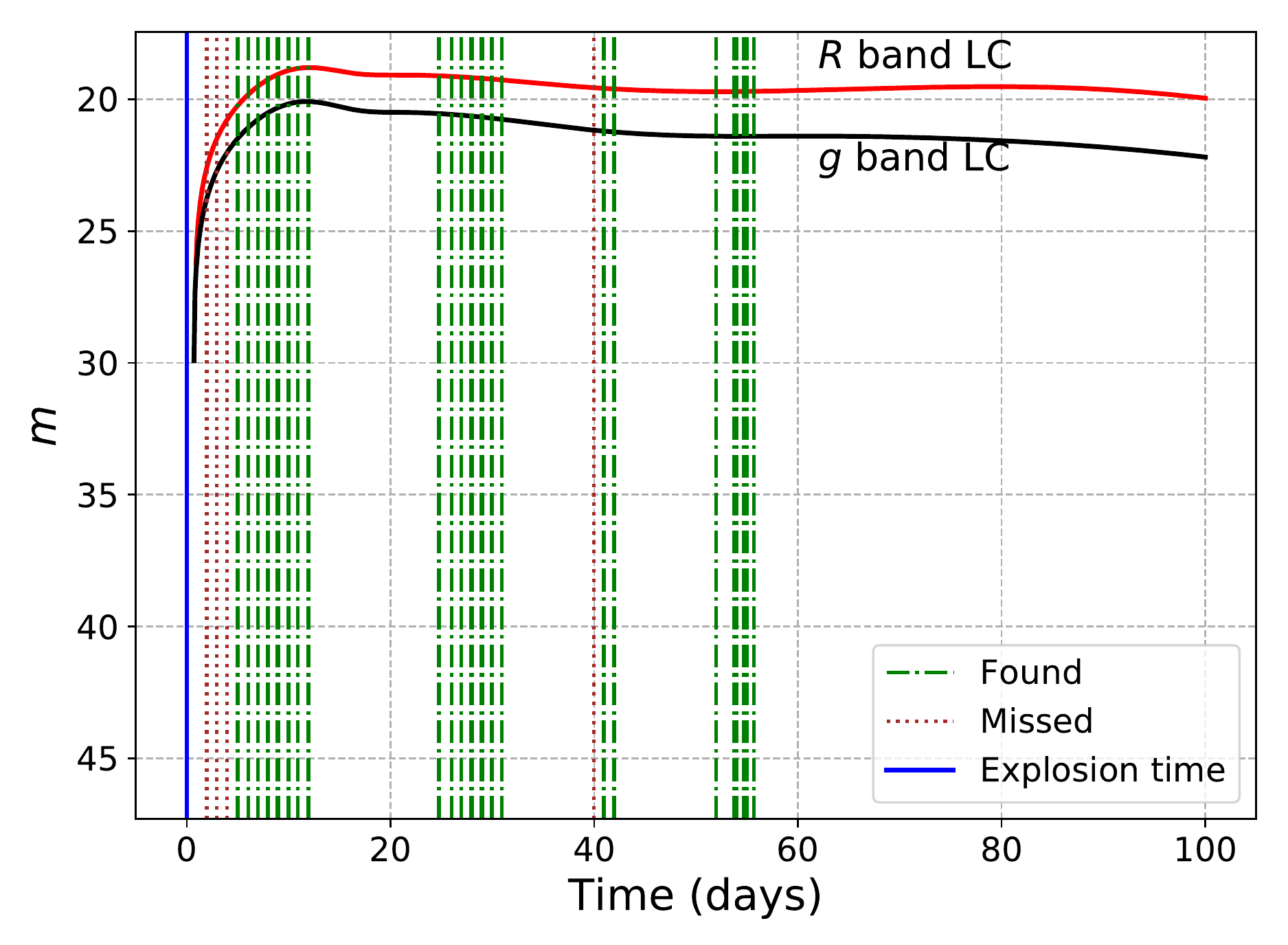}
    \end{center}
    \caption{\textbf{Upper panel}: An example of a SN~IIp lightcurve,
    with the apparent magnitude, $m$ on the y-axis and time on x-axis.
    The lightcurve is shown in the iPTF $R$ and $g$ bands. The
    observations of the telescope are shown as vertical lines.
    \textbf{Lower panel}: The same lightcurve is plotted, however,
    the vertical lines now represent the recovery by single epoch classifier.
    One can identify the only $g$ band observation (around 40 days) being
    missed due to fainter magnitude in the $g$ band.}
    \label{fig:rates_example_sn2p_lightcurve}
\end{figure}

In contrast to the well-defined Ia lightcurves with their typical timescales of
several weeks, we also wanted to explore longer-timescale lightcurves as a limiting
case. Therefore, we consider type IIp supernovae and compute their space-time sensitive
volume in similar lines as Sec.~\ref{subsec:lightcurve_ensemble}.
In general, type II supernovae (SNe~II) vary in lightcurve morphology and are categorized
in various subtypes \citep{li_2011}. Specifically, type IIp lightcurves have a distinct
``plateau'' feature after the rise lasting for about 100 days after explosion, as shown
in Fig.~\ref{fig:rates_example_sn2p_lightcurve}.  The intrinsic brightness,
$M_B \sim -16.75$, is significantly lower than that of SNe~Ia \citep{richardson_2014}.
Hence, we expect the space-time sensitive volume to be lower than that
of the SNe~Ia. When considering the Ia lightcurves in Sec.~\ref{subsec:salt2_lightcurves},
the SALT2 model parameters were used to tune possible lightcurve morphologies.
Here we take a simpler approach and consider a time-series model from
\cite{1999ApJ...521...30G} (named \texttt{nugent-sn2p} in the \texttt{sncosmo}
package) to compute the flux up to 100 days from the explosion time.
Thus, while simulating the SNe~IIp in space-time, the only change to the
lightcurve shape is the ``stretch'' depending on the cosmological redshift.

We simulate $\sim 9.1 \times 10^5$ SN~IIp lightcurves uniform in sky location, observer
time and co-moving volume up to a redshift, $z = 0.1$. Like the SNe~Ia, each SN~IIp is assigned
a host galaxy surface brightness from the surface brightness distribution of galaxies
in iPTF and a $E(B-V)$ extinction value from IIp extinction distribution in \cite{hatano}.
In this case, we use the criteria that the lightcurve must be recovered a minimum of five
epochs, brighter than 20th magnitude in a span of 3 weeks within the 100 days post explosion.
The iPTF observing schedule along with the single-epoch classifier is used to compute the
detectability in each epoch. We obtain the result:
\begin{equation}
    \langle VT \rangle_{\mathrm{IIp}} = 7.80 \pm 0.76 \times 10^{-4}
    \; \mathrm{Gpc^{3}\,yr},
\label{eq:rates_sn2p_VT_value}
\end{equation}
where the error includes the statistical error from the Monte-Carlo integration
and the $6\%$ systematic uncertainty from the single-epoch classifier (see
Sec.~\ref{sec:multi_dim_efficiency}). The median sensitive redshift is
found to be $z_\text{median}^{\text{IIp}} = 0.038$.

%% file: discussion.tex
In this work, we provide a methodology to assess the transient
detectability taking into account the intrinsic transient
properties and the observing conditions of fast cadence transient
surveys. This is done by injecting fake point source transients
into the images, running image subtraction on them and finding out
the parameter space where they are found by the image subtraction
pipeline.
The joint detectability is evaluated using the machinery
of supervised machine learning trained on the missed and found
fake transients. This step mimics the action of the image subtraction
pipeline at every epoch and forms the single-epoch detectability.
Consequently, the lightcurve morphology and the survey observing schedule
is used to compute the space-time volume sensitivity of particular
transients. We consider the case of the intermediate Palomar Transient
Factory (iPTF) and evaluate the single-epoch detectability and then
use its observing schedule to compute the space-time volume sensitivity
of type Ia supernovae (SNe~Ia). We also do a preliminary analysis of
type IIp supernovae (SNe~IIp). Note that the space-time volume sensitivity
could be computed for any general transient, using its lightcurve morphology;
SN~Ia or IIp is an example. In the case of SNe~Ia, the remaining piece in the
estimation of the volumetric rate is a systematic number count to be
obtained via an archival search into iPTF data.
While we defer this to a future work, we outline our plan of action here.

\subsection{Rates}
The computation of the rate posterior assumes the likelihood
of observing $N$ candidate events is an inhomogeneous Poisson process
\citep{loredo95, farr15}. Our \emph{search} will filter the SN~Ia population
based on the model presented in Sec.~\ref{sec:rates} at the expense of some
contamination from other transient types, potentially with similar lightcurve
morphologies. If the mean count of these impurities is $\Lambda_0$, the
likelihood function is:
\begin{eqnarray}
    p\left(N \vert \Lambda_0, \lambdasne\right) \propto
        \left(\Lambda_0 p_0 + \lambdasne \psne\right)^N
        \nonumber \\
        \times \exp{\left(- \Lambda_0 - \lambdasne\right)},
\label{eq:discussion_likelihood}
\end{eqnarray}
where $\psne$ ($p_0$) is the \emph{a priori} weight that a
transient is (isn't) a SN~Ia after the filtering process.
With a suitable choice of prior, we can use Bayes' theorem to obtain
the posterior. Considering the Jeffreys' prior:
\begin{equation}
    p\left(\Lambda_0, \lambdasne\right) = \frac{1}{\sqrt{\Lambda_0}}
                                          \frac{1}{\sqrt{\lambdasne}},
\label{eq:discussion_prior}
\end{equation}
the posterior takes the form:
\begin{eqnarray}
    p\left(\Lambda_0, \lambdasne\right \vert N) &\propto &
        p\left(N \vert \Lambda_0, \lambdasne\right)
        p\left(\Lambda_0, \lambdasne\right) \nonumber \\
    &\propto &
    \frac{\left(\Lambda_0 p_0 + \lambdasne \psne\right)^N}
         {\sqrt{\Lambda_0\lambdasne}} \nonumber \\
    &&\times \exp{\left(- \Lambda_0 - \lambdasne\right)}.
\label{eq:discussion_posterior_full}
\end{eqnarray}
Integrating out the nuisance parameter, $\Lambda_0$, we have the
marginalized posterior on
$\lambdasne = R\langle VT \rangle$, or equivalently on $R$:
\begin{eqnarray}
    p\left(R\vert N\right) &=&
        \int_0^{\infty} p\left(\Lambda_0, \lambdasne\right \vert N)
        \mathrm{d}\Lambda_0 \nonumber \\
    &\propto & \frac{e^{-R\langle VT \rangle}}
               {\sqrt{R\langle VT \rangle}} \times
     \left[\left(R\langle VT \rangle \psne\right)^N + \right. \nonumber \\
    &&\left. \frac{N}{2}p_0\left(R\langle VT \rangle \psne\right)^{N-1}\right],
\label{eq:discussion_posterior}
\end{eqnarray}
where we expand Eq.~(\ref{eq:discussion_posterior_full}) and integrate, keeping
terms up to linear order in $p_0$ since we expect that $p_0 \ll \psne$.

\subsection{Approximate SN~Ia Count in iPTF}
Type Ia supernova rates have been studied earlier in the literature
\citep{Dilday_2008, subaru, 10.1093/mnras/stz258}.
Deep field instruments have provided estimates of the Ia rate out to high
redshift \citep{subaru}. The intermediate Palomar Transient Factory, being
an all sky survey has a comparatively lower sensitivity to SNe~Ia at
\replaced{$z_{\mathrm{median}} = 0.098$}{$z_{\text{median}}^{\text{Ia}} = 0.099$},
evaluated in Sec.~\ref{sec:rates}. The SDSS-II
supernova survey has estimated the volumetric SN~Ia rate at $z\approx 0.1$ to be
$R_{\mathrm{SNIa}}^{\mathrm{SDSS-II}}\sim 2.9^{+1.07}_{-0.75} \times 10^{-5}
\mathrm{Mpc^{-3}yr^{-1}}$ \citep{Dilday_2008}. Using our estimate of the
space-time sensitive volume from Eq.~(\ref{eq:rates_VT_value}), an estimate of the
count of SNe~Ia in iPTF is $630 - 1160$. This is consistent with $1035$ objects
tagged ``SN~Ia'' during the survey time.

\subsection{Future Work}
While the number of transients tagged as ``SN~Ia'' by human scanners during
iPTF survey time seem consistent with our ballpark above, the systematic
uncertainty of such a classification remains unquantified. The quantities $p_0$,
$\psne$ and $N$ in Eq.~(\ref{eq:discussion_posterior}) require a
systematic search into the iPTF archival data to retrieve the candidate count
and systematic errors associated with such a classification.
We defer this and the computation of SN~Ia volumetric rate to a future
work in the series.

The methodology developed here facilitates the computation of space-time
volume sensitivities of general transient types. Of particular interest
are the fast transients in iPTF archival data as discussed in \cite{ho18}.
Also, the observation of the ``kilonova'' resulting from the binary neutron
star merger, GW170817 \citep{2017ApJ...848L..12A, dynamical_ejecta, Abbott_2017},
hints towards the association of transients to binary neutron star
mergers. There is no evidence of detection of such a transient in the iPTF data,
in which case rate upper limits could be placed due to non-detection.

%% file: acknowledgements.tex
This work was supported by GROWTH (Global Relay of Observatories Watching
Transients Happen) project under the National Science Foundation (NSF) grant no.
1545949. The research used resources of the National Energy Research Scientific
Computing Center (NERSC), a DOE Office of Science User Facility supported by the
Office of Science of the U.S. Department of Energy under Contract No. DE-AC02-05CH11231.
D.C. acknowledges the use of computing facilities provided
by NERSC and by Leonard E. Parker Center for Gravitation, Cosmology and Astrophysics
at University of Wisconsin-Milwaukee. The latter is supported by NSF Awards PHY-1626190
and PHY-1607585. P.E.N. acknowledges support from the DOE through DE-FOA-0001088, Analytical Modeling
for Extreme-Scale Computing Environments. D.C. would like to thank Shaon Ghosh, Jolien
Creighton, Siddharth Mohite, Angela Van Sistine and Lin Yan for helpful discussions.
\added{
We thank the anonymous referee for helpful comments.
}

\software{
SExtractor \citep{sextractor}, HOTPANTS \citep{2015ascl.soft04004B},
Astropy \citep{2018AJ....156..123A},
sncosmo \citep{sncosmo}, scikit-learn \citep{scikit-learn}, 
Matplotlib \citep{Hunter:2007}, scipy \citep{scipy}, numpy \citep{numpy},
pandas \citep{mckinney-proc-scipy-2010}, jupyter (\url{https://jupyter.org/}),
SQLAlchemy (\url{https://www.sqlalchemy.org/}).
}

%% file: appendix.tex
\section{Classifier Single-epoch performance}\label{sec:appendix}
\begin{figure*}
\begin{center}
    \subfloat{
        \includegraphics[width=0.45\textwidth]
        {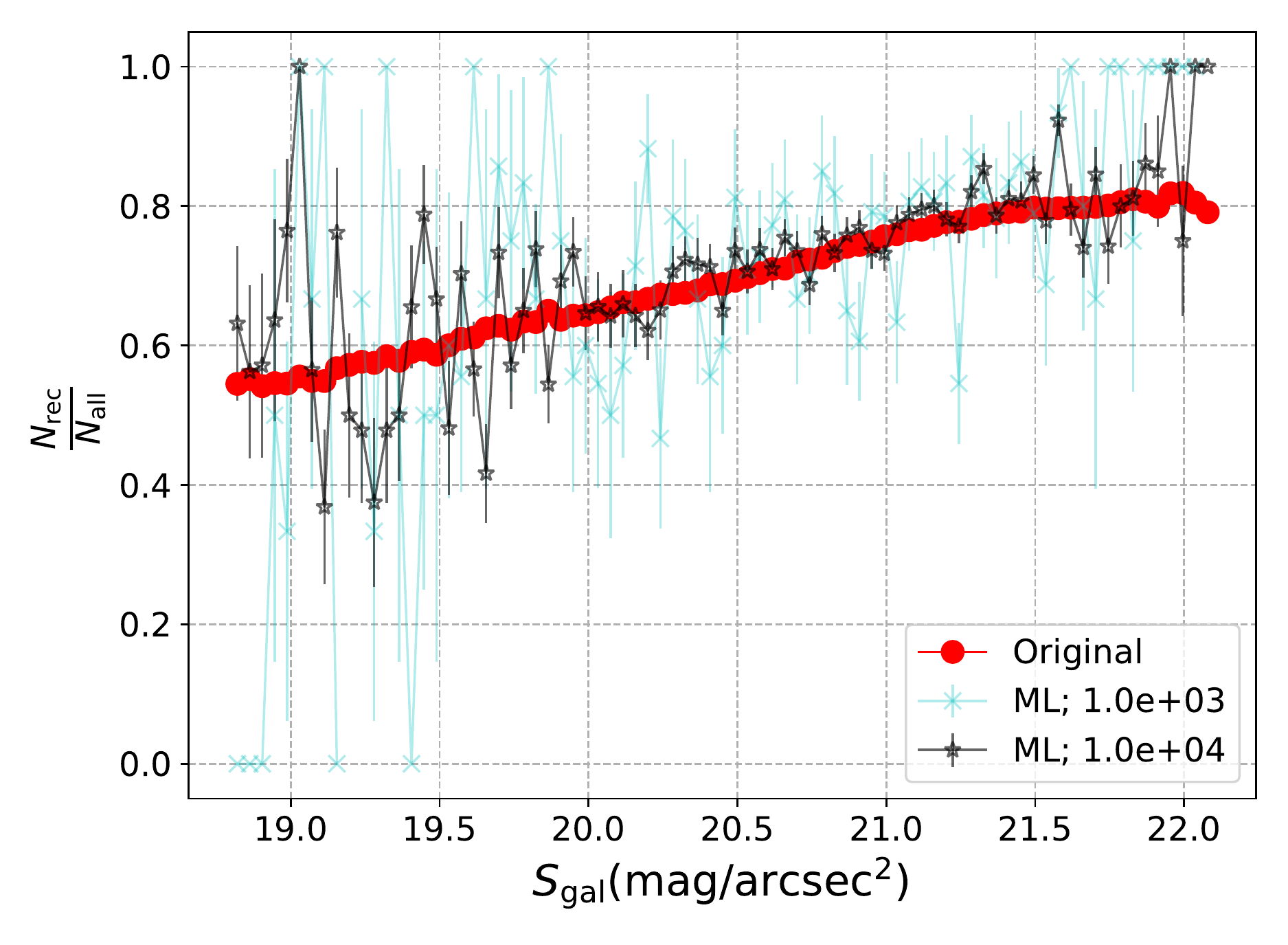}
    }
    \subfloat{    
        \includegraphics[width=0.45\textwidth]
        {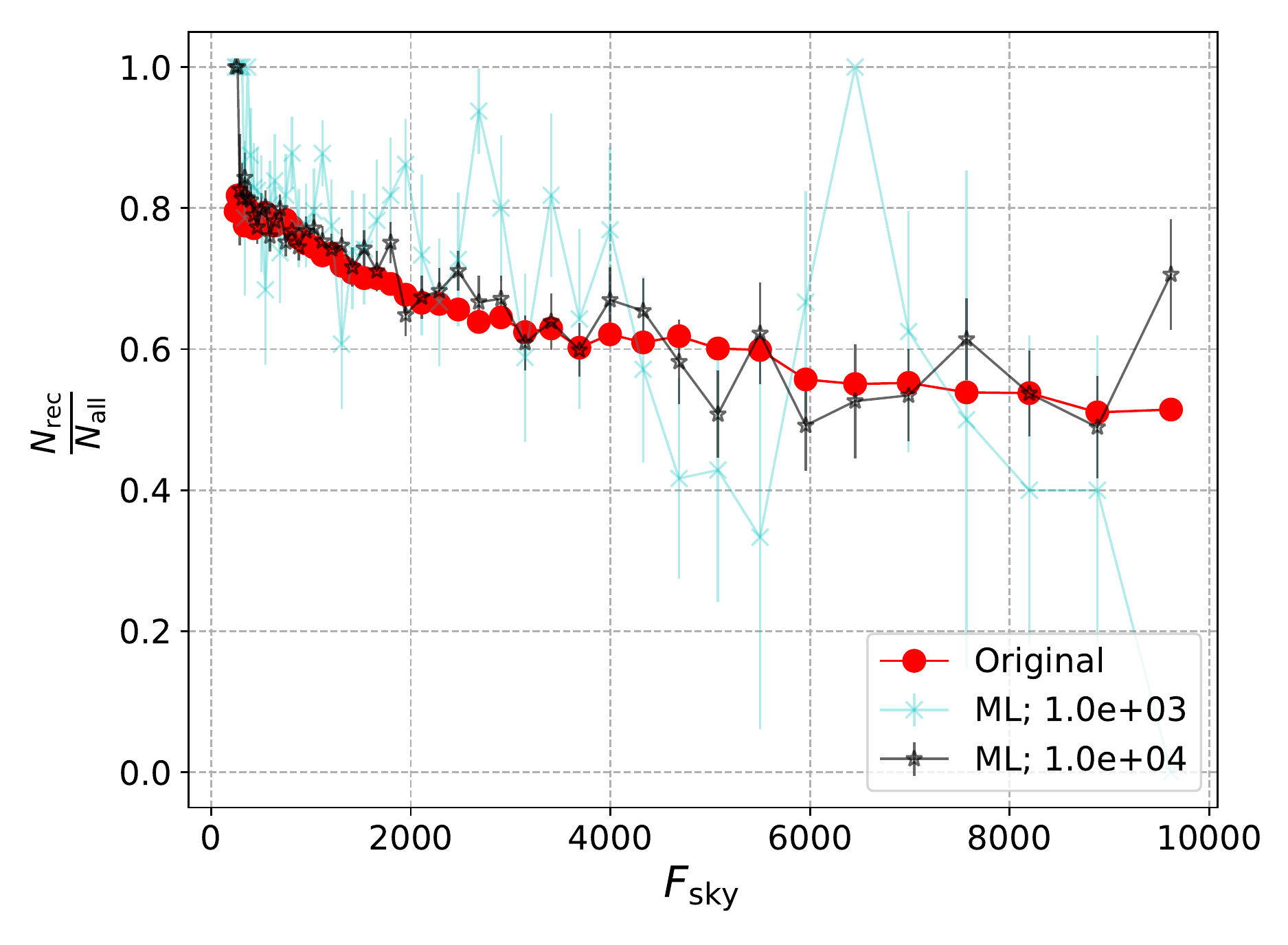}
    } \\
    \subfloat{
        \includegraphics[width=0.45\textwidth]
        {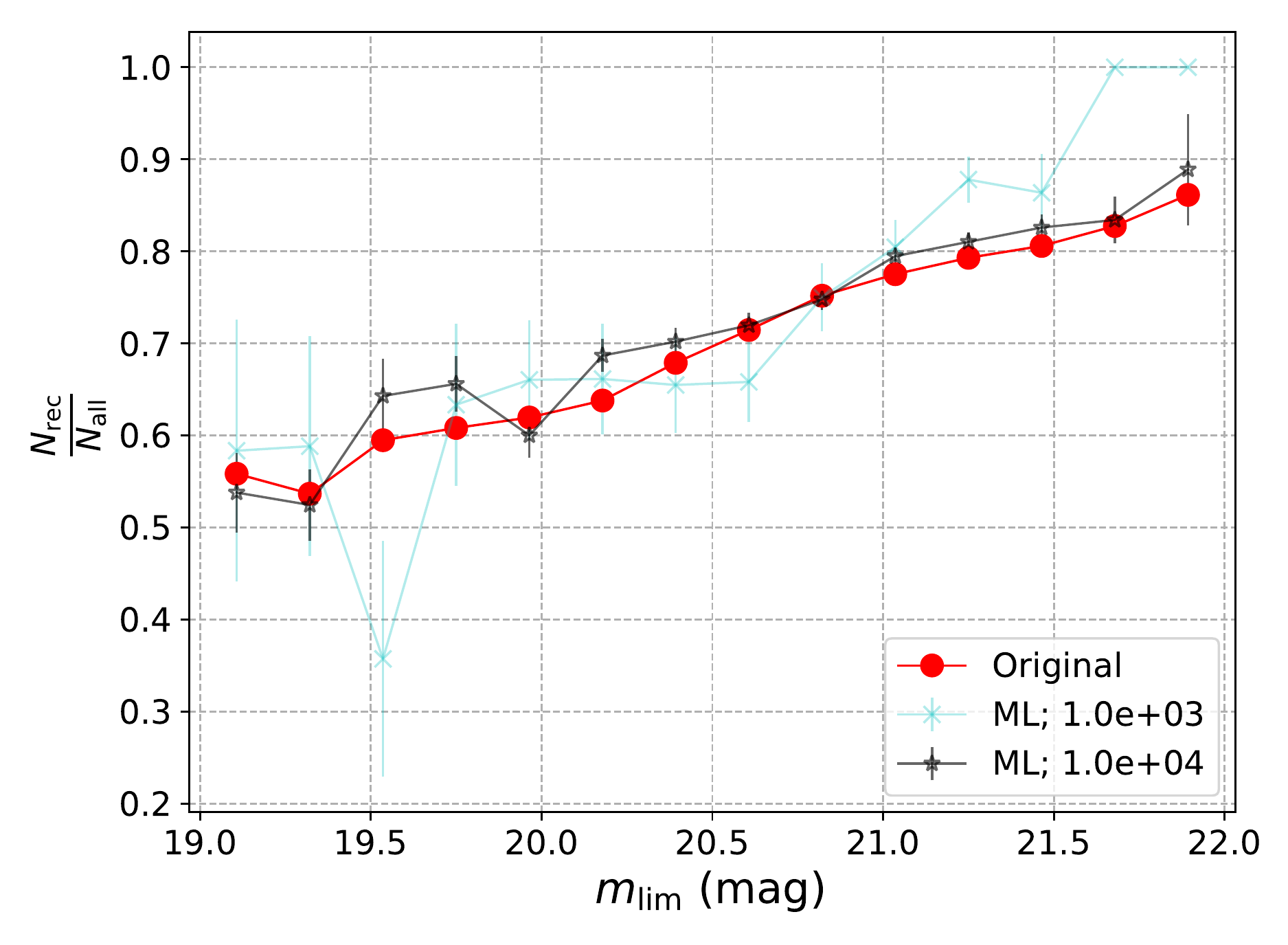}
    }
    \subfloat{
        \includegraphics[width=0.45\textwidth]
        {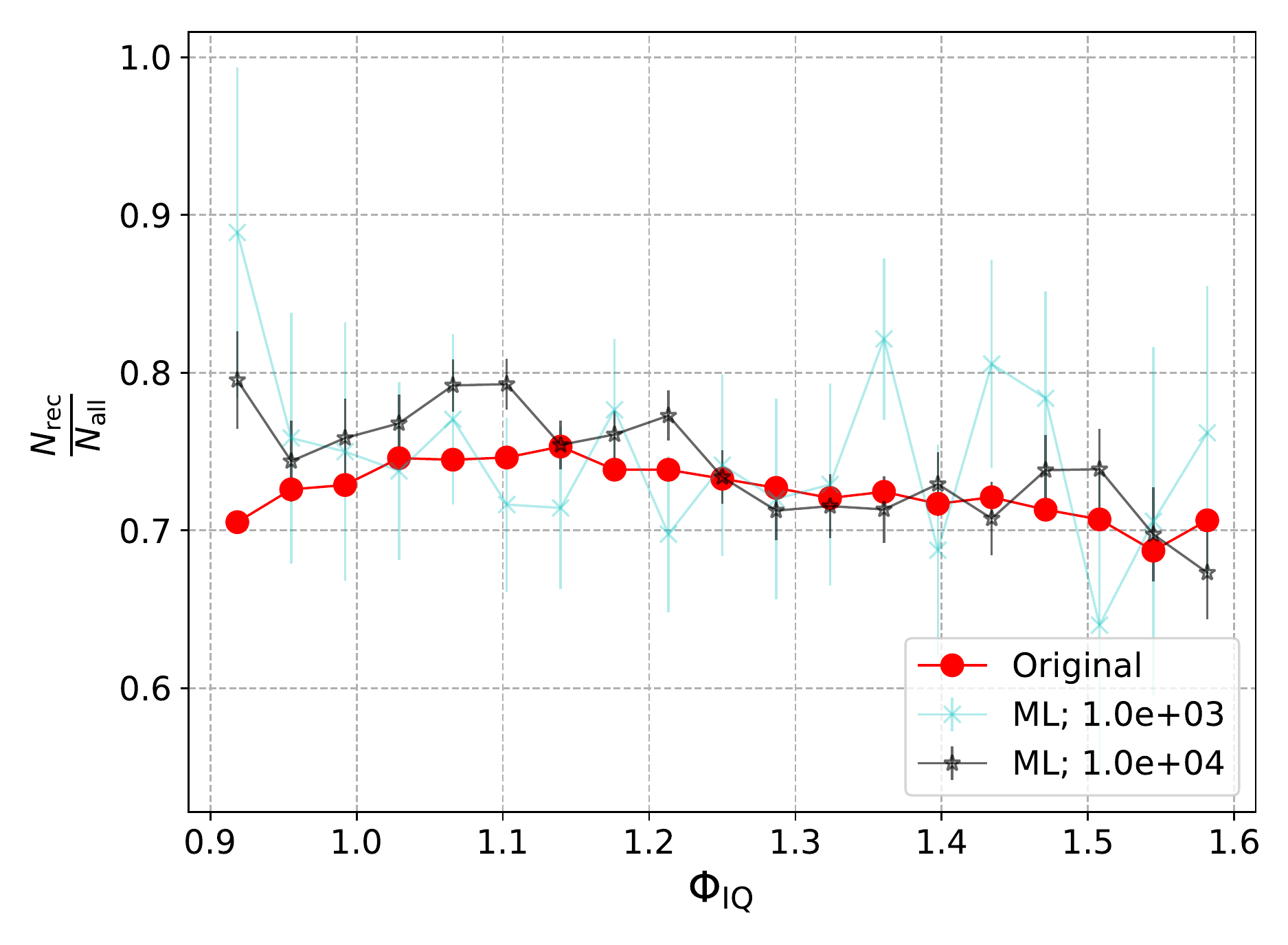}
    }
\end{center}
\caption{This figure is an extension of Fig.~\ref{fig:efficiency_ML_performance}. We
compare the performance of the marginalized single parameter efficiency of the trained
classifier compared to that of the original distributions in Fig.~\ref{fig:efficiency_single_parameter}.
We see the behavior of the ISP being reproduced by feeding the classifier a few thousand points.
}
\label{fig:ml_performace_extended}
\end{figure*}
In Fig.~\ref{fig:efficiency_ML_performance}, we made a comparison
between the marginalized single parameter efficiency for the single-epoch
transient brightness from the classifier predictions.
Here, we show it for the remaining parameters. While the final classifier is
trained on the full dataset, to make the comparison, we train it on
$90\%$ of the total fake point source simulations we performed, as mentioned
in Sec.~\ref{sec:point_source_transients}. From the remaining $10\%$ sample size,
we make a random selection of points (progressively increasing), feed them to the
classifier and bin the results in the same manner as in
Fig.~\ref{fig:efficiency_single_parameter} to compare marginalized efficiency plots.
These are shown in Fig.~\ref{fig:ml_performace_extended} and
Fig.~\ref{fig:efficiency_ML_performance}, the latter presented earlier.
We see that the behavior starts to converge to that of the ISP in a few
thousand points.
\FloatBarrier

%% file: main.bbl
\begin{thebibliography}{}
\expandafter\ifx\csname natexlab\endcsname\relax\def\natexlab#1{#1}\fi

\bibitem[{Abbott {et~al.}(2017{\natexlab{a}})}]{dynamical_ejecta}
Abbott, B.~P., {et~al.} 2017{\natexlab{a}}, The Astrophysical Journal, 850, L39

\bibitem[{Abbott {et~al.}(2017{\natexlab{b}})}]{Abbott_2017}
---. 2017{\natexlab{b}}, The Astrophysical Journal, 848, L13

\bibitem[{{Abbott} {et~al.}(2017){Abbott}, {Abbott}, {Abbott}, {Acernese},
  {Ackley}, {Adams}, {Adams}, {Addesso}, {Adhikari}, {Adya}, \&
  et~al.}]{2017ApJ...848L..12A}
{Abbott}, B.~P., {Abbott}, R., {Abbott}, T.~D., {et~al.} 2017, ApJ Lett., 848,
  L12

\bibitem[{{Astropy ~Collaboration} {et~al.}(2018){Astropy ~Collaboration},
  {Price-Whelan}, {Sip{\H o}cz}, {G{\"u}nther}, {Lim}, {Crawford}, {Conseil},
  {Shupe}, {Craig}, {Dencheva}, {Ginsburg}, {VanderPlas}, {Bradley},
  {P{\'e}rez-Su{\'a}rez}, {de Val-Borro}, {Aldcroft}, {Cruz}, {Robitaille},
  {Tollerud}, {Ardelean}, {Babej}, {Bach}, {Bachetti}, {Bakanov}, {Bamford},
  {Barentsen}, {Barmby}, {Baumbach}, {Berry}, {Biscani}, {Boquien}, {Bostroem},
  {Bouma}, {Brammer}, {Bray}, {Breytenbach}, {Buddelmeijer}, {Burke},
  {Calderone}, {Cano Rodr{\'{\i}}guez}, {Cara}, {Cardoso}, {Cheedella},
  {Copin}, {Corrales}, {Crichton}, {D'Avella}, {Deil}, {Depagne}, {Dietrich},
  {Donath}, {Droettboom}, {Earl}, {Erben}, {Fabbro}, {Ferreira}, {Finethy},
  {Fox}, {Garrison}, {Gibbons}, {Goldstein}, {Gommers}, {Greco}, {Greenfield},
  {Groener}, {Grollier}, {Hagen}, {Hirst}, {Homeier}, {Horton}, {Hosseinzadeh},
  {Hu}, {Hunkeler}, {Ivezi{\'c}}, {Jain}, {Jenness}, {Kanarek}, {Kendrew},
  {Kern}, {Kerzendorf}, {Khvalko}, {King}, {Kirkby}, {Kulkarni}, {Kumar},
  {Lee}, {Lenz}, {Littlefair}, {Ma}, {Macleod}, {Mastropietro}, {McCully},
  {Montagnac}, {Morris}, {Mueller}, {Mumford}, {Muna}, {Murphy}, {Nelson},
  {Nguyen}, {Ninan}, {N{\"o}the}, {Ogaz}, {Oh}, {Parejko}, {Parley}, {Pascual},
  {Patil}, {Patil}, {Plunkett}, {Prochaska}, {Rastogi}, {Reddy Janga},
  {Sabater}, {Sakurikar}, {Seifert}, {Sherbert}, {Sherwood-Taylor}, {Shih},
  {Sick}, {Silbiger}, {Singanamalla}, {Singer}, {Sladen}, {Sooley},
  {Sornarajah}, {Streicher}, {Teuben}, {Thomas}, {Tremblay}, {Turner},
  {Terr{\'o}n}, {van Kerkwijk}, {de la Vega}, {Watkins}, {Weaver}, {Whitmore},
  {Woillez}, {Zabalza}, \& {Astropy Contributors}}]{2018AJ....156..123A}
{Astropy ~Collaboration}, {Price-Whelan}, A.~M., {Sip{\H o}cz}, B.~M., {et~al.}
  2018, \aj, 156, 123

\bibitem[{Barbary(2014)}]{sncosmo}
Barbary, K. 2014, doi:{10.5281/zenodo.11938}

\bibitem[{{Becker}(2015)}]{2015ascl.soft04004B}
{Becker}, A. 2015, {HOTPANTS: High Order Transform of PSF ANd Template
  Subtraction}, , , ascl:1504.004

\bibitem[{{Bertin, E.} \& {Arnouts, S.}(1996)}]{sextractor}
{Bertin, E.}, \& {Arnouts, S.} 1996, Astron. Astrophys. Suppl. Ser., 117, 393

\bibitem[{{Betoule} {et~al.}(2014){Betoule}, {Kessler}, {Guy}, {Mosher},
  {Hardin}, {Biswas}, {Astier}, {El-Hage}, {Konig}, {Kuhlmann}, {Marriner},
  {Pain}, {Regnault}, {Balland}, {Bassett}, {Brown}, {Campbell}, {Carlberg},
  {Cellier-Holzem}, {Cinabro}, {Conley}, {D'Andrea}, {DePoy}, {Doi}, {Ellis},
  {Fabbro}, {Filippenko}, {Foley}, {Frieman}, {Fouchez}, {Galbany}, {Goobar},
  {Gupta}, {Hill}, {Hlozek}, {Hogan}, {Hook}, {Howell}, {Jha}, {Le Guillou},
  {Leloudas}, {Lidman}, {Marshall}, {M{\"o}ller}, {Mour{\~a}o}, {Neveu},
  {Nichol}, {Olmstead}, {Palanque-Delabrouille}, {Perlmutter}, {Prieto},
  {Pritchet}, {Richmond}, {Riess}, {Ruhlmann-Kleider}, {Sako}, {Schahmaneche},
  {Schneider}, {Smith}, {Sollerman}, {Sullivan}, {Walton}, \& {Wheeler}}]{jla}
{Betoule}, M., {Kessler}, R., {Guy}, J., {et~al.} 2014, \aap, 568, A22

\bibitem[{Bloom {et~al.}(2013)Bloom, Brink, Richards, Rice, Wainwright,
  Poznanski, \& Negahban}]{real_bogus_2}
Bloom, J.~S., Brink, H., Richards, J.~W., {et~al.} 2013, Monthly Notices of the
  Royal Astronomical Society, 435, 1047

\bibitem[{Brown {et~al.}(2019)Brown, Kochanek, Stanek, Thompson, Beacom,
  Bersier, Brimacombe, Holoien, Shappee, Prieto, Chen, Dong, \&
  Stritzinger}]{10.1093/mnras/stz258}
Brown, J.~S., Kochanek, C.~S., Stanek, K.~Z., {et~al.} 2019, Monthly Notices of
  the Royal Astronomical Society, 484, 3785

\bibitem[{Cao {et~al.}(2016)Cao, Nugent, \& Kasliwal}]{cao2016}
Cao, Y., Nugent, P.~E., \& Kasliwal, M.~M. 2016, Publications of the
  Astronomical Society of the Pacific, 128, 114502

\bibitem[{Dilday {et~al.}(2008)Dilday, Kessler, Frieman, Holtzman, Marriner,
  Miknaitis, Nichol, Romani, Sako, Bassett, Becker, Cinabro, DeJongh, Depoy,
  Doi, Garnavich, Hogan, Jha, Konishi, Lampeitl, Marshall, McGinnis, Prieto,
  Riess, Richmond, Schneider, Smith, Takanashi, Tokita, van~der Heyden, Yasuda,
  Zheng, Barentine, Brewington, Choi, Crotts, Dembicky, Harvanek, Im,
  Ketzeback, Kleinman, Krzesi{\'{n}}ski, Long, Malanushenko, Malanushenko,
  McMillan, Nitta, Pan, Saurage, Snedden, Watters, Wheeler, \&
  York}]{Dilday_2008}
Dilday, B., Kessler, R., Frieman, J.~A., {et~al.} 2008, The Astrophysical
  Journal, 682, 262

\bibitem[{Drake {et~al.}(2009)Drake, Djorgovski, Mahabal, Beshore, Larson,
  Graham, Williams, Christensen, Catelan, Boattini, Gibbs, Hill, \&
  Kowalski}]{catalina}
Drake, A.~J., Djorgovski, S.~G., Mahabal, A., {et~al.} 2009, The Astrophysical
  Journal, 696, 870

\bibitem[{{Farr} {et~al.}(2015){Farr}, {Gair}, {Mandel}, \& {Cutler}}]{farr15}
{Farr}, W.~M., {Gair}, J.~R., {Mandel}, I., \& {Cutler}, C. 2015, \prd, 91,
  023005

\bibitem[{Fitzpatrick(1999)}]{f99dust}
Fitzpatrick, E.~L. 1999, Publications of the Astronomical Society of the
  Pacific, 111, 63

\bibitem[{Frohmaier {et~al.}(2018)Frohmaier, Sullivan, Maguire, \&
  Nugent}]{Frohmaier_2018}
Frohmaier, C., Sullivan, M., Maguire, K., \& Nugent, P. 2018, The Astrophysical
  Journal, 858, 50

\bibitem[{Frohmaier {et~al.}(2017)Frohmaier, Sullivan, Nugent, Goldstein, \&
  DeRose}]{frohmaier_2017}
Frohmaier, C., Sullivan, M., Nugent, P.~E., Goldstein, D.~A., \& DeRose, J.
  2017, The Astrophysical Journal Supplement Series, 230, 4

\bibitem[{Gal-Yam {et~al.}(2007)Gal-Yam, Filippenko, Jannuzi, Maoz, Schweiker,
  Silverman, Sharon, Doi, Fukugita, Yasuda, Foley, Morokuma, Oda, Totani, \&
  Poznanski}]{subaru}
Gal-Yam, A., Filippenko, A.~V., Jannuzi, B.~T., {et~al.} 2007, Monthly Notices
  of the Royal Astronomical Society, 382, 1169

\bibitem[{{Gilliland} {et~al.}(1999){Gilliland}, {Nugent}, \&
  {Phillips}}]{1999ApJ...521...30G}
{Gilliland}, R.~L., {Nugent}, P.~E., \& {Phillips}, M.~M. 1999, \apj, 521, 30

\bibitem[{Guy {et~al.}(2007)Guy, Astier, Baumont, Hardin, Pain, Regnault, Basa,
  Carlberg, Conley, Fabbro, {et~al.}}]{salt2}
Guy, J., Astier, P., Baumont, S., {et~al.} 2007, Astronomy \& Astrophysics,
  466, 11

\bibitem[{Hatano {et~al.}(1998)Hatano, Branch, \& Deaton}]{hatano}
Hatano, K., Branch, D., \& Deaton, J. 1998, The Astrophysical Journal, 502, 177

\bibitem[{{Hinshaw} {et~al.}(2013){Hinshaw}, {Larson}, {Komatsu}, {Spergel},
  {Bennett}, {Dunkley}, {Nolta}, {Halpern}, {Hill}, \&
  {Odegard}}]{2013ApJS..208...19H}
{Hinshaw}, G., {Larson}, D., {Komatsu}, E., {et~al.} 2013, \apjs, 208, 19

\bibitem[{Ho {et~al.}(2018)Ho, Kulkarni, Nugent, Zhao, Rusu, Cenko, Ravi,
  Kasliwal, Perley, Adams, Bellm, Brady, Fremling, Gal-Yam, Kann, Kaplan,
  Laher, Masci, Ofek, Sollerman, \& Urban}]{ho18}
Ho, A. Y.~Q., Kulkarni, S.~R., Nugent, P.~E., {et~al.} 2018, The Astrophysical
  Journal, 854, L13

\bibitem[{{Holoien} {et~al.}(2019){Holoien}, {Brown}, {Vallely}, {Stanek},
  {Kochanek}, {Shappee}, {Prieto}, {Dong}, {Brimacombe}, {Bishop}, {Bose},
  {Beacom}, {Bersier}, {Chen}, {Chomiuk}, {Falco}, {Holmbo}, {Jayasinghe},
  {Morrell}, {Pojmanski}, {Shields}, {Strader}, {Stritzinger}, {Thompson},
  {Wo{\'z}niak}, {Bock}, {Cacella}, {Carballo}, {Cruz}, {Conseil}, {Farfan},
  {Fernandez}, {Kiyota}, {Koff}, {Krannich}, {Marples}, {Masi}, {Monard},
  {Mu{\~n}oz}, {Nicholls}, {Post}, {Stone}, {Trappett}, \&
  {Wiethoff}}]{2019MNRAS.484.1899H}
{Holoien}, T.~W.-S., {Brown}, J.~S., {Vallely}, P.~J., {et~al.} 2019, \mnras,
  484, 1899

\bibitem[{Hunter(2007)}]{Hunter:2007}
Hunter, J.~D. 2007, Computing In Science \& Engineering, 9, 90

\bibitem[{{Ivezi{\'c}} {et~al.}(2008){Ivezi{\'c}}, {Kahn}, {Tyson}, {Abel},
  {Acosta}, {Allsman}, {Alonso}, {AlSayyad}, {Anderson}, {Andrew}, {Angel},
  {Angeli}, {Ansari}, {Antilogus}, {Araujo}, {Armstrong}, {Arndt}, {Astier},
  {Aubourg}, {Auza}, {Axelrod}, {Bard}, {Barr}, {Barrau}, {Bartlett}, {Bauer},
  {Bauman}, {Baumont}, {Becker}, {Becla}, {Beldica}, {Bellavia}, {Bianco},
  {Biswas}, {Blanc}, {Blazek}, {Blandford}, {Bloom}, {Bogart}, {Bond},
  {Borgland}, {Borne}, {Bosch}, {Boutigny}, {Brackett}, {Bradshaw}, {Nielsen
  Brand t}, {Brown}, {Bullock}, {Burchat}, {Burke}, {Cagnoli}, {Calabrese},
  {Callahan}, {Callen}, {Chandrasekharan}, {Charles-Emerson}, {Chesley},
  {Cheu}, {Chiang}, {Chiang}, {Chirino}, {Chow}, {Ciardi}, {Claver},
  {Cohen-Tanugi}, {Cockrum}, {Coles}, {Connolly}, {Cook}, {Cooray}, {Covey},
  {Cribbs}, {Cui}, {Cutri}, {Daly}, {Daniel}, {Daruich}, {Daubard}, {Daues},
  {Dawson}, {Delgado}, {Dellapenna}, {de Peyster}, {de Val-Borro}, {Digel},
  {Doherty}, {Dubois}, {Dubois-Felsmann}, {Durech}, {Economou}, {Eracleous},
  {Ferguson}, {Figueroa}, {Fisher-Levine}, {Focke}, {Foss}, {Frank}, {Freemon},
  {Gangler}, {Gawiser}, {Geary}, {Gee}, {Geha}, {Gessner}, {Gibson}, {Gilmore},
  {Glanzman}, {Glick}, {Goldina}, {Goldstein}, {Goodenow}, {Graham},
  {Gressler}, {Gris}, {Guy}, {Guyonnet}, {Haller}, {Harris}, {Hascall},
  {Haupt}, {Hernand ez}, {Herrmann}, {Hileman}, {Hoblitt}, {Hodgson}, {Hogan},
  {Huang}, {Huffer}, {Ingraham}, {Innes}, {Jacoby}, {Jain}, {Jammes}, {Jee},
  {Jenness}, {Jernigan}, {Jevremovi{\'c}}, {Johns}, {Johnson}, {Johnson},
  {Jones}, {Juramy-Gilles}, {Juri{\'c}}, {Kalirai}, {Kallivayalil}, {Kalmbach},
  {Kantor}, {Karst}, {Kasliwal}, {Kelly}, {Kessler}, {Kinnison}, {Kirkby},
  {Knox}, {Kotov}, {Krabbendam}, {Krughoff}, {Kub{\'a}nek}, {Kuczewski},
  {Kulkarni}, {Ku}, {Kurita}, {Lage}, {Lambert}, {Lange}, {Langton}, {Le
  Guillou}, {Levine}, {Liang}, {Lim}, {Lintott}, {Long}, {Lopez}, {Lotz},
  {Lupton}, {Lust}, {MacArthur}, {Mahabal}, {Mand elbaum}, {Marsh}, {Marshall},
  {Marshall}, {May}, {McKercher}, {McQueen}, {Meyers}, {Migliore}, {Miller},
  {Mills}, {Miraval}, {Moeyens}, {Monet}, {Moniez}, {Monkewitz}, {Montgomery},
  {Mueller}, {Muller}, {Mu{\~n}oz Arancibia}, {Neill}, {Newbry}, {Nief},
  {Nomerotski}, {Nordby}, {O'Connor}, {Oliver}, {Olivier}, {Olsen},
  {O'Mullane}, {Ortiz}, {Osier}, {Owen}, {Pain}, {Palecek}, {Parejko},
  {Parsons}, {Pease}, {Peterson}, {Peterson}, {Petravick}, {Libby Petrick},
  {Petry}, {Pierfederici}, {Pietrowicz}, {Pike}, {Pinto}, {Plante}, {Plate},
  {Price}, {Prouza}, {Radeka}, {Rajagopal}, {Rasmussen}, {Regnault}, {Reil},
  {Reiss}, {Reuter}, {Ridgway}, {Riot}, {Ritz}, {Robinson}, {Roby}, {Roodman},
  {Rosing}, {Roucelle}, {Rumore}, {Russo}, {Saha}, {Sassolas}, {Schalk},
  {Schellart}, {Schindler}, {Schmidt}, {Schneider}, {Schneider}, {Schoening},
  {Schumacher}, {Schwamb}, {Sebag}, {Selvy}, {Sembroski}, {Seppala}, {Serio},
  {Serrano}, {Shaw}, {Shipsey}, {Sick}, {Silvestri}, {Slater}, {Smith},
  {Smith}, {Sobhani}, {Soldahl}, {Storrie-Lombardi}, {Stover}, {Strauss},
  {Street}, {Stubbs}, {Sullivan}, {Sweeney}, {Swinbank}, {Szalay}, {Takacs},
  {Tether}, {Thaler}, {Thayer}, {Thomas}, {Thukral}, {Tice}, {Trilling},
  {Turri}, {Van Berg}, {Vand en Berk}, {Vetter}, {Virieux}, {Vucina}, {Wahl},
  {Walkowicz}, {Walsh}, {Walter}, {Wang}, {Wang}, {Warner}, {Wiecha},
  {Willman}, {Winters}, {Wittman}, {Wolff}, {Wood-Vasey}, {Wu}, {Xin},
  {Yoachim}, {Zhan}, \& {for the LSST Collaboration}}]{lsst}
{Ivezi{\'c}}, {\v{Z}}., {Kahn}, S.~M., {Tyson}, J.~A., {et~al.} 2008, arXiv
  e-prints, arXiv:0805.2366

\bibitem[{Jones {et~al.}(2001--)Jones, Oliphant, Peterson, {et~al.}}]{scipy}
Jones, E., Oliphant, T., Peterson, P., {et~al.} 2001--, {SciPy}: Open source
  scientific tools for {Python}, , , [Online; accessed <today>]

\bibitem[{Kaiser {et~al.}(2010)Kaiser, Burgett, Chambers, Denneau, Heasley,
  Jedicke, Magnier, Morgan, Onaka, \& Tonry}]{panstarrs1}
Kaiser, N., Burgett, W., Chambers, K., {et~al.} 2010, The Pan-STARRS wide-field
  optical/NIR imaging survey, , , doi:10.1117/12.859188

\bibitem[{{Kulkarni}(2016)}]{ztf_kulkarni}
{Kulkarni}, S.~R. 2016, in American Astronomical Society Meeting Abstracts,
  Vol. 227, American Astronomical Society Meeting Abstracts \#227, 314.01

\bibitem[{Law {et~al.}(2009)Law, Kulkarni, Dekany, Ofek, Quimby, Nugent,
  Surace, Grillmair, Bloom, Kasliwal, Bildsten, Brown, Cenko, Ciardi, Croner,
  Djorgovski, van Eyken, Filippenko, Fox, Gal-Yam, Hale, Hamam, Helou, Henning,
  Howell, Jacobsen, Laher, Mattingly, McKenna, Pickles, Poznanski, Rahmer, Rau,
  Rosing, Shara, Smith, Starr, Sullivan, Velur, Walters, \& Zolkower}]{ptf}
Law, N.~M., Kulkarni, S.~R., Dekany, R.~G., {et~al.} 2009, Publications of the
  Astronomical Society of the Pacific, 121, 1395

\bibitem[{{Li} {et~al.}(2011){Li}, {Chornock}, {Leaman}, {Filippenko},
  {Poznanski}, {Wang}, {Ganeshalingam}, \& {Mannucci}}]{li_2011}
{Li}, W., {Chornock}, R., {Leaman}, J., {et~al.} 2011, \mnras, 412, 1473

\bibitem[{{Loredo} \& {Wasserman}(1995)}]{loredo95}
{Loredo}, T.~J., \& {Wasserman}, I.~M. 1995, \apjs, 96, 261

\bibitem[{{Maoz} \& {Mannucci}(2012)}]{2012PASA...29..447M}
{Maoz}, D., \& {Mannucci}, F. 2012, \pasa, 29, 447

\bibitem[{McKinney(2010)}]{mckinney-proc-scipy-2010}
McKinney, W. 2010, in Proceedings of the 9th Python in Science Conference, ed.
  S.~van~der Walt \& J.~Millman, 51 -- 56

\bibitem[{Nugent {et~al.}(2015)Nugent, Cao, \& Kasliwal}]{nugent}
Nugent, P., Cao, Y., \& Kasliwal, M. 2015, The Palomar transient factory, , ,
  doi:10.1117/12.2085383

\bibitem[{Pedregosa {et~al.}(2011)Pedregosa, Varoquaux, Gramfort, Michel,
  Thirion, Grisel, Blondel, Prettenhofer, Weiss, Dubourg, Vanderplas, Passos,
  Cournapeau, Brucher, Perrot, \& Duchesnay}]{scikit-learn}
Pedregosa, F., Varoquaux, G., Gramfort, A., {et~al.} 2011, Journal of Machine
  Learning Research, 12, 2825

\bibitem[{{Richardson} {et~al.}(2014){Richardson}, {Jenkins}, {Wright}, \&
  {Maddox}}]{richardson_2014}
{Richardson}, D., {Jenkins}, Robert~L., I., {Wright}, J., \& {Maddox}, L. 2014,
  \aj, 147, 118

\bibitem[{Sako {et~al.}(2007)Sako, Bassett, Becker, Cinabro, DeJongh, Depoy,
  Dilday, Doi, Frieman, Garnavich, Hogan, Holtzman, Jha, Kessler, Konishi,
  Lampeitl, Marriner, Miknaitis, Nichol, Prieto, Riess, Richmond, Romani,
  Schneider, Smith, SubbaRao, Takanashi, Tokita, van~der Heyden, Yasuda, Zheng,
  Barentine, Brewington, Choi, Dembicky, Harnavek, Ihara, Im, Ketzeback,
  Kleinman, Krzesi{\'{n}}ski, Long, Malanushenko, Malanushenko, McMillan,
  Morokuma, Nitta, Pan, Saurage, \& Snedden}]{sdss}
Sako, M., Bassett, B., Becker, A., {et~al.} 2007, The Astronomical Journal,
  135, 348

\bibitem[{{Shanks} {et~al.}(2015){Shanks}, {Metcalfe}, {Chehade}, {Findlay},
  {Irwin}, {Gonzalez-Solares}, {Lewis}, {Yoldas}, {Mann}, {Read}, {Sutorius},
  \& {Voutsinas}}]{atlas}
{Shanks}, T., {Metcalfe}, N., {Chehade}, B., {et~al.} 2015, \mnras, 451, 4238

\bibitem[{{van der Walt} {et~al.}(2011){van der Walt}, {Colbert}, \&
  {Varoquaux}}]{numpy}
{van der Walt}, S., {Colbert}, S.~C., \& {Varoquaux}, G. 2011, Computing in
  Science Engineering, 13, 22

\end{thebibliography}
